# Designing Stress-Adaptive Dense Suspensions using Dynamic Covalent Chemistry

Grayson L. Jackson,[a,*] Joseph M. Dennis,[b] Neil D. Dolinski,[c] Michael van der Naald,[a,d] Hojin Kim,[a,c] Christopher Eom,[c] Stuart J. Rowan,[c,e,f] and Heinrich M. Jaeger[a,d]

[a] James Franck Institute, University of Chicago, 929 E. 57th Street, Chicago, Illinois 60637, United States

[b] Combat Capabilities and Development Command, Army Research Laboratory, Aberdeen Proving Ground, MD 21005, USA

[c] Pritzker School of Molecular Engineering, University of Chicago, 5640 S. Ellis Avenue, Chicago, Illinois 60637, United States

[d] Department of Physics, University of Chicago, 5720 S. Ellis Avenue, Chicago, Illinois 60637, United States

[e] Department of Chemistry, University of Chicago, 5735 S Ellis Avenue, Chicago, Illinois 60637, United States

[f] Chemical and Engineering Sciences Division, Argonne National Laboratory, 9700 Cass Avenue, Lemont, Illinois 60439, United States

**ABSTRACT:** The non-Newtonian behaviors of dense suspensions are central to their use in technological and industrial applications and arise from a network of particle-particle contacts that dynamically adapts to imposed shear. Reported herein are studies aimed at exploring how dynamic covalent chemistry between particles and the polymeric solvent can be used to tailor such stress-adaptive contact networks leading to their unusual rheological behaviors. Specifically, a room temperature dynamic thia-Michael bond is employed to rationally tune the equilibrium constant ($K_{eq}$) of the polymeric solvent to the particle interface. It is demonstrated that low $K_{eq}$ leads to shear thinning while high $K_{eq}$ produces antithixotropy, a rare phenomenon where the viscosity increases with shearing time. It is proposed that an increase in $K_{eq}$ increases the polymer graft density at the particle surface and that antithixotropy primarily arises from partial debonding of the polymeric graft/solvent from the particle surface and the formation of polymer bridges between particles. Thus, the implementation of dynamic covalent chemistry provides a new molecular handle with which to tailor the macroscopic rheology of suspensions by introducing programable time dependence. These studies open the door to energy absorbing materials that not only sense mechanical inputs and adjust their dissipation as a function of time or shear rate but can switch between these two modalities on demand.

## INTRODUCTION

Concentrated or "dense" suspensions of particles in a Newtonian suspending solvent can autonomously sense applied stress and adapt their mechanical properties in response. This feature makes them promising for smart applications in additive manufacturing,[1-3] coatings and lubrication,[4, 5] vibration dampening,[6] and impact mitigation.[7, 8] These highly loaded polymer nanocomposites commonly exhibit non-Newtonian flow behaviors, where the viscosity depends on the shear rate or shear stress and may also depend on shearing time. A suspension is termed shear thinning or shear thickening when the viscosity decreases or increases with shear rate, respectively.[9, 10] Additionally, the suspension can be thixotropic if viscosity decreases over time at a fixed shear rate or antithixotropic (*also termed negative thixotropic or rheopectic*) if it increases over time at a fixed shear rate.[11] In particular, antithixotropic materials are attractive because they can transform from liquid-like to solid-like under shear and the resulting properties of the shear-induced gel such as yield stress[12] and conductivity[13] can be programmed using their shear history. The ability to controllably tune viscosity (or dissipation) with time would also be relevant to vibration dampening[14] and impact mitigation. However, antithixotropic materials are rare and few design rules exist to target this unique and useful non-Newtonian behavior.

The non-Newtonian rheology of dense suspensions originates from microscopic constraints on interparticle motion.[5, 10, 15-21] Shear thinning arises from *stress-released* constraints which are broken upon increasing shear rate or stress, whereas *stress-activated* constraints (formed by increasing shear rate or stress) cause shear thickening. While the macroscopic viscosity in either case is constant at a given shear rate or stress, constraints are constantly breaking and reforming within a structurally dynamic network of non-covalent interparticle contacts. When the viscosity does evolve with time at a constant shear rate as in thixotropy (or antithixotropy), this signals the release (or formation) of constraints.[11] While great strides have been made in understanding the constraint-based physics of dense suspensions, an emerging challenge is to connect microscopic constraints to specific chemical interactions.[22]

To this end, prior work has focused on manipulating non-covalent chemical interactions such as van der Waals forces, solvation forces,[23-25] depletion attraction,[26, 27] steric stabilization,[28-31] and hydrogen bonding.[32-35] These seminal studies provide a conceptual framework to rationalize basic shear thickening or shear thinning behavior in terms of the relative strength of particle-particle and particle-solvent interactions (*i.e.* solubility).[24, 36] If particle-particle attractions cannot be overcome by particle-solvent interactions, then the suspension possesses adhesive constraints at rest which are broken by shear (*shear thinning*). If the particle-solvent interaction strength is increased[24, 25, 36, 37] and the particle-particle attraction is diminished, by using surfactants[28, 29] or a covalently grafted steric barrier,[30, 31] then particles are dispersed at rest, yet can form frictionally stabilized contacts under shear (*shear thickening*). The understanding from this prior work was established using simple non-covalent interactions, yet synthetic organic chemistry offers nearly limitless potential to tune particle-particle and particle-solvent interactions, thus presenting a new frontier for designing responsive dense suspensions.

Dynamic covalent chemistry (DCC) has recently emerged as a method to engineer stress-adaptive functional polymeric materials.[38-44] Like the non-covalent interactions described above, dynamic covalent bonds are able to dissociate and re-associate under equilibrium conditions, though typically they require a catalyst or external stimulus to access this reversibility.[42] When integrated into a dynamic covalent network (DCN) or covalent adaptable network (CAN), dynamic bonds enable structural reorganization under mechanical stress. This behavior is quite sensitive to the dynamic equilibrium constant ($K_{eq}$).[39, 45] Past work also used interfacial DCC to integrate functionalized filler particles into crosslinked resins and studied the stress relaxation of these nanocomposites, though the DCCs used required exogenous catalysts.[46-50] As opposed to these nanocomposite systems with a crosslinked suspending matrix, dense suspensions possess a fluid matrix which allows particle migration and interparticle constraints to be formed or released under shear. While there have been studies of nanoparticle gels stabilized by dynamic covalent crosslinks,[51-53] these works were primarily concerned with self-assembly rather than shear rheology. Within the context of a dense suspension, an ideal DCC would allow ambient temperature dynamic exchange without a catalyst as well as a readily tunable bond strength ($K_{eq}$), both of which are achievable using a special class of thia-Michael (tM) reactions.

The tM reaction is the addition of a thiol to a thia-Michael accepting (tMA) electron-poor olefin to form a thioether adduct.[54] As demonstrated by foundational small molecule studies,[55-57] selection of certain substituents adjacent to the double bond can result in catalyst-free dynamic tM bonds at ambient temperatures. Examples of this include benzalcyanoacetate (BCA) and benzalcyanoacetamide (BCAm)-based tMAs. An advantage of BCA or BCAm-based tMAs is that the $K_{eq}$ can be tuned by varying the electron-donating/withdrawing nature of the R-substituents attached to the phenyl ring, which has been exploited in dynamic polymer networks,[45] adhesives[58] and hydrogels.[59, 60]

Taking advantage of the room temperature, catalyst-free, dynamic covalent bonds of the tM reaction with BCAm tMAs, presented here is the first study aimed at exploring the use of DCC in dense suspensions. Specifically, thiol-coated particles are dispersed in a fluid matrix comprised of a low molecular weight BCAm end-capped polymer. Importantly, this polymeric tMA solvent can form dynamic tM bonds at the particle surface to yield a dynamic brush layer, (Figure 1A) with a bonding strength ($K_{eq}$) orders of magnitude larger than is achievable through a single hydrogen bond. While conventional non-covalent dense suspensions (with only hydrogen bonding interactions at the interface) (**NCSs**) exhibit shear thinning, these dynamic covalent suspensions (**DCSs**) exhibit antithixotropy wherein the viscosity reversibly increases under shear and relaxes upon shear cessation. Interestingly, the rheology of **DCSs** can be tuned between shear thinning and antithixotropy by varying $K_{eq}$ of the tM bond, which in turn affects the dynamic graft density at the particle surface. Moreover, studies of monotopic BCAm tMAs reveal that antithixotropy in **DCSs** arises from partial debonding of the particle grafts from the surface under shear and formation of polymer bridges between particles.

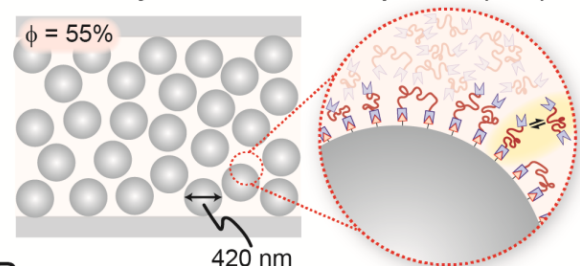

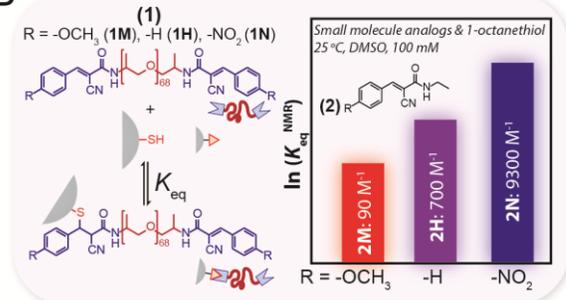

**Figure 1.** (*A*) Illustration depicting a dense dynamic covalent suspension (**DCS**). These high volume fraction ($\phi = 55\%$) **DCSs** contain particles which can form room temperature dynamic covalent bonds with the surrounding fluid polymer matrix, resulting in a bonded polymer graft layer which exchanges dynamically (*inset*). (*B*) Realization of **DCSs** using dynamic covalent thia-Michael (tM) chemistry. Chemical structure of ditopic poly(propylene glycol) benzalcyanoacetamide (BCAm) thia-Michael acceptors (tMAs) (**1**) with different R-substituents at the *para*-position of the $\beta$-phenyl ring, R = -OCH$_3$ (**1M**), -H (**1H**), -NO$_2$ (**1N**). These tMAs can form dynamic tM bonds at the surface of thiol-functionalized particles which exchange dynamically under ambient conditions without a catalyst. Small molecule analogs (**2M**, **2H**, or **2N**) were used to assess how temperature and chemistry affect the dynamic equilibrium constant ($K_{eq}^{NMR}$).

## RESULTS AND DISCUSSION

**Material Synthesis and DCS Preparation.** To experimentally realize the concept in Figure 1A, thiol-coated particles were prepared by grafting (3-mercaptopropyl)trimethoxysilane onto commercially available silica particles using literature procedures (Scheme S1).[61] After surface functionalization, the particle diameter was 417 ± 30 nm (Figure S1) and the particle surface was covered with 0.5 thiols/nm$^2$ as determined by NMR.[61]

The tMA end-capped polymer was synthesized in 2 steps from an amine terminated $M_n \sim 4000$ g/mol poly(propylene glycol) (PPG) core. Acid-catalyzed condensation with cyanoacetic acid and subsequent Knoevenagel condensation with different benzaldehydes were used to synthesize 3 ditopic BCAm polymers (Schemes S2-3 and Table S1). These are referred to by their R-substituents at the *para*-position of the $\beta$-phenyl ring:

methoxy (R= -OCH₃) (**1M**), unsubstituted (R = -H) (**1H**), nitro (R = -NO₂) (**1N**) (Figure 1B).

To understand the baseline effect of temperature and R-substituent on the dynamic equilibrium constant ($K_{eq}$), small molecule analogs **2M**, **2H**, **2N** were synthesized and NMR was used to measure $K_{eq}^{NMR}$ under 100 mM equimolar (with 100 mM 1-octanethiol) conditions in DMSO-$d_6$ over the temperature range 25 – 77 °C (Schemes S4-5, Figures S2-7, and Table S2). These experiments show that the value of $K_{eq}^{NMR}$ can be tuned by over a factor of ~$10^3$ using temperature and chemistry: $K_{eq}^{NMR}$ decreases upon heating and, at a constant temperature, shows a trend of **2N** > **2H** > **2M** (Figure 1B). As the reaction proceeds through a charged enolate intermediate,[57] reaction rates and overall equilibrium are expected to be impacted by solvent polarity. In addition to this, $K_{eq}$ for **DCSs** represents polymeric tMA binding to surface thiols, and the entropic penalty for polymer chain stretching is not accounted for by $K_{eq}^{NMR}$ and would lead to a lower effective $K_{eq}$ in **DCSs**. With respect to these points, the small molecule controls are treated as estimates, however, the trends in temperature and chemistry from the model studies are expected to translate to **DCSs**.

**DCSs** at a particle volume fraction (φ) of 55% were prepared with either **1M**, **1H**, or **1N** as the suspending solvent to yield **DCS-1M**, **DCS-1H**, or **DCS-1N**. A control non-covalent dense suspension (**NCS-OH**) was prepared at the same φ and with the same particles but with 4000 g/mol hydroxy-terminated PPG as the polymer matrix. From φ, particle density, and thiol surface coverage, it is estimated that [-SH] ~ 0.016 M and [tMA] ~ 0.46 M in the liquid phase of these suspensions, a nearly 30-fold excess of the tMA. In other words, the surface thiol group is the limiting reagent and leads to a high bonding fraction and subsequent polymer grafting at the particle surface. The tM adducts are envisioned to serve as a dynamic brush layer that depends on the dynamic bond strength ($K_{eq}$) with the remaining unbound tMAs serving as the carrier fluid (Figure 1A).

**NCS and DCS Rheology.** NCS-OH serves as a useful starting point to understand **DCS** rheology. **NCS-OH** exhibits conventional shear thinning where the viscosity decreases with increasing shear rate (γ̇) (Figure 2A, black trace). In this figure, the reduced viscosity $\eta_r$ is used to isolate the viscosity contribution of the particles from that of the suspending polymeric solvent (Figure S8). The forward (increasing) and backwards (decreasing) shear rate ramps overlay quite well for **NCS-OH**, illustrating no processing hysteresis and a viscosity which is independent of shearing time. As explored by others,[19, 24, 36] shear thinning behavior in systems like **NCS-OH** can be understood as a solubility mismatch wherein hydrogen bonding or van der Waals forces between particle and solvent are not strong enough to overcome interparticle attractions. This leads to a stress-bearing (i.e. high viscosity) network of adhesive particle-particle contacts at rest which is disrupted by shear and leads to shear thinning.[15, 29] These stress-released interparticle contacts rapidly reform upon shear cessation and the suspension viscosity does not strongly depend on the shear history.

In stark contrast to the conventional shear thinning of **NCS-OH**, introduction of dynamic tM chemistry at the particle surface in **DCS-1M** leads to rich time-dependent rheology (Figure 2A). The measured viscosity on the forward shear rate ramp is low, while it is much higher upon the backwards shear rate ramp, indicating a strong hysteresis (Figure 2A, inset). However, the higher viscosity state decays upon shear cessation. Increasing the waiting time at each point in the shear rate ramp

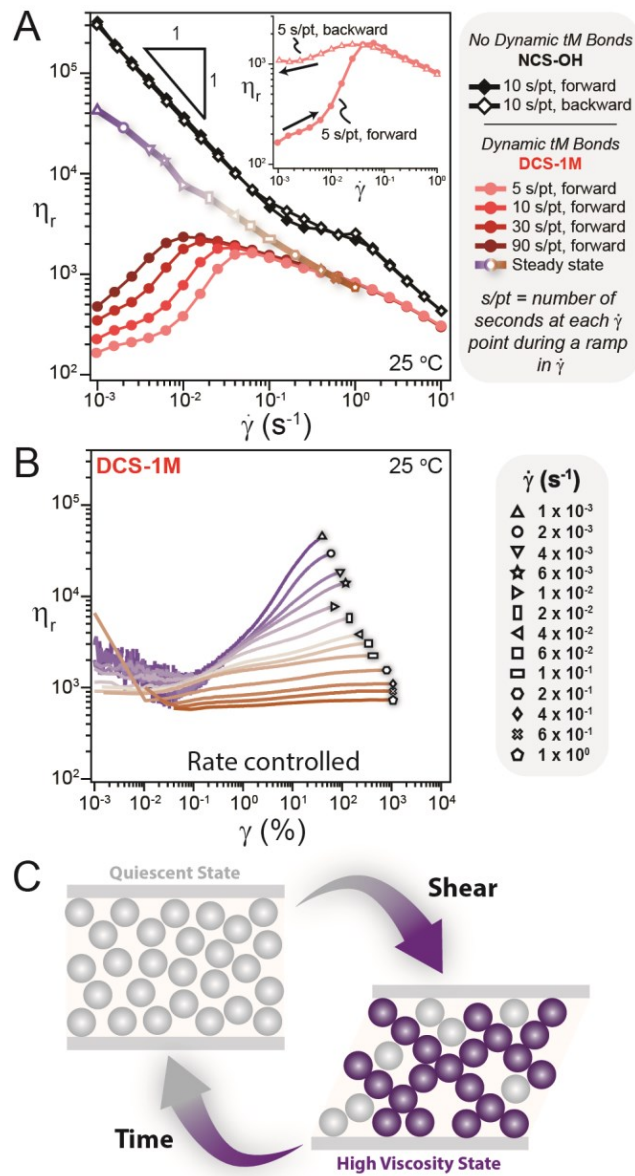

**Figure 2.** (*A*) Suspensions with only non-covalent hydrogen bonding interactions between particles and solvent (**NCS-OH**) exhibit reversible shear thinning. In contrast, dynamic covalent suspensions such as **DCS-1M** exhibit antithixotropy. Reduced viscosity $\eta_r$ versus shear rate (γ̇) for **DCS-1M** with different waiting times (s/pt is the number of seconds at each γ̇ point during a ramp in γ̇) reveal that $\eta_r$ increases as a function of shearing time and eventually approaches a steady-state. Similarly, comparison of the increasing (*forward*) and decreasing (*backward*) γ̇ ramps reveal hysteresis (*inset*). (*B*) Evolution of $\eta_r$ as a function of strain (γ) at a constant γ̇ to reach a steady state. (*C*) Schematic depiction of antithixotropy, wherein shear reversibly transforms a low viscosity quiescent state into a higher viscosity state through the formation of a stress-bearing particle network (*purple*).

leads to a higher measured viscosity at low shear rates while the data were identical at higher shear rates (Figure 2A). At a constant shear rate, the viscosity evolves as a function of strain (γ) and shows an initial decay from the pre-sheared state followed by growth and a plateau at a high strain value (Figure 2B). The high strain value, which is interpreted as an approximate "steady state" viscosity plotted in Figure 2A, decreases with increasing shear rate or exhibits steady state shear thinning behavior. In line with the constant shear rate measurements, constant stress (*creep*) measurements of **DCS-1M** reveal a viscosity bifurcation:[12, 62] the viscosity diverges for $\sigma \leq 10$ Pa and flows for $\sigma \geq 100$ Pa (Figure S9). In other words, **DCS-1M** exhibits a yield stress, but only under shear. Qualitatively similar behavior was observed in constant shear rate and creep measurements of **DCS-1H** (Figure S10) and **DCS-1N** (Figure S11). This reversible increase in viscosity indicates that these **DCSs** are antithixotropic, *i.e.*, the opposite of more conventional thixotropy, where shear forms a stress-bearing particle network which returns to its equilibrium quiescent state upon shear cessation (Figure 2C).[4, 11]

Small-amplitude oscillatory shear (SAOS) was used to track the decay of the complex viscosity ($\eta^*$) to understand how these shear-induced structures in **DCSs** relax upon shear cessation (Figure S12). These SAOS experiments were conducted immediately following the constant shear rate experiments shown in Figures 2B and S10-11 (see Figure S12 for experimental protocol). **DCSs** subjected to lower shear rates, which typically had larger $\eta_r$ plateau values prior to SAOS, exhibited a slower decay of $\eta^*$. In other words, more robust particle contact networks prior to shear cessation typically persisted longer once oscillatory shear was applied. It is worth pointing out that for shear-induced networks with similar $\eta_r$, the decay of $\eta^*$ does not clearly correlate with dynamic bond $K_{eq}$ but does coincide with the trend in the viscosity of the dynamic tMA oil matrix (generally the slowest for **DCS-1N**, followed by **DCS-1M**, and then **DCS-1H**). This observation suggests that polymer diffusion also plays a role in the relaxation process.

**Tuning Macroscopic Rheology with $K_{eq}$.** As indicated by the small molecule studies (Figure 1B and S7), $K_{eq}$ and the dynamic brush layer density in **DCSs** can be systematically varied using temperature and chemistry. As such, temperature dependent rheology of **DCSs** was performed over the range $0 - 80$ °C and shear rate ramps were used to identify antithixotropy or lack thereof (Figures 3A, S13-14). As expected, **NCS-OH** exhibits reversible shear thinning or mild thixotropy over the entire temperature range (Figures 3A and S13). However, the **DCSs** are much more sensitive to temperature and changes in $K_{eq}$. While **DCS-1M** exhibits antithixotropy at 10 °C (Figure S14) and 30 °C, decreasing $K_{eq}$ by heating $\geq 40$ °C leads to strong and reversible shear thinning with a slope of nearly -1 on a log-log plot of $\eta_r$ vs. $\dot{\gamma}$ (Figure 3A). At 70 °C it is apparent that $\eta_r$ at a given shear rate is significantly larger even than that reached in the steady state at 25 °C. In other words, the stress-bearing particle network formed at rest at 70 °C is much stronger than that formed under shear at 25 °C. A similar transition from antithixotropy to shear thinning with mild thixotropy is observed when heating **DCS-1H** but the transition occurs when heating between 60 and 70 °C (Figure 3A). The area of the hysteresis loop generally decreases upon heating **DCS-1M** or **DCS-1H**, which could reflect a gradual transition between antithixotropy and shear thinning. The larger $K_{eq}$ for **DCS-1N** leads to antithixotropy over the entire investigated range with an increasing mismatch between the forward and backward shear ramps. This trend for **DCS-1N** is particularly obvious at 70 °C where the viscosity at the end of the hysteresis loop is 3 orders of magnitude larger than its initial value.

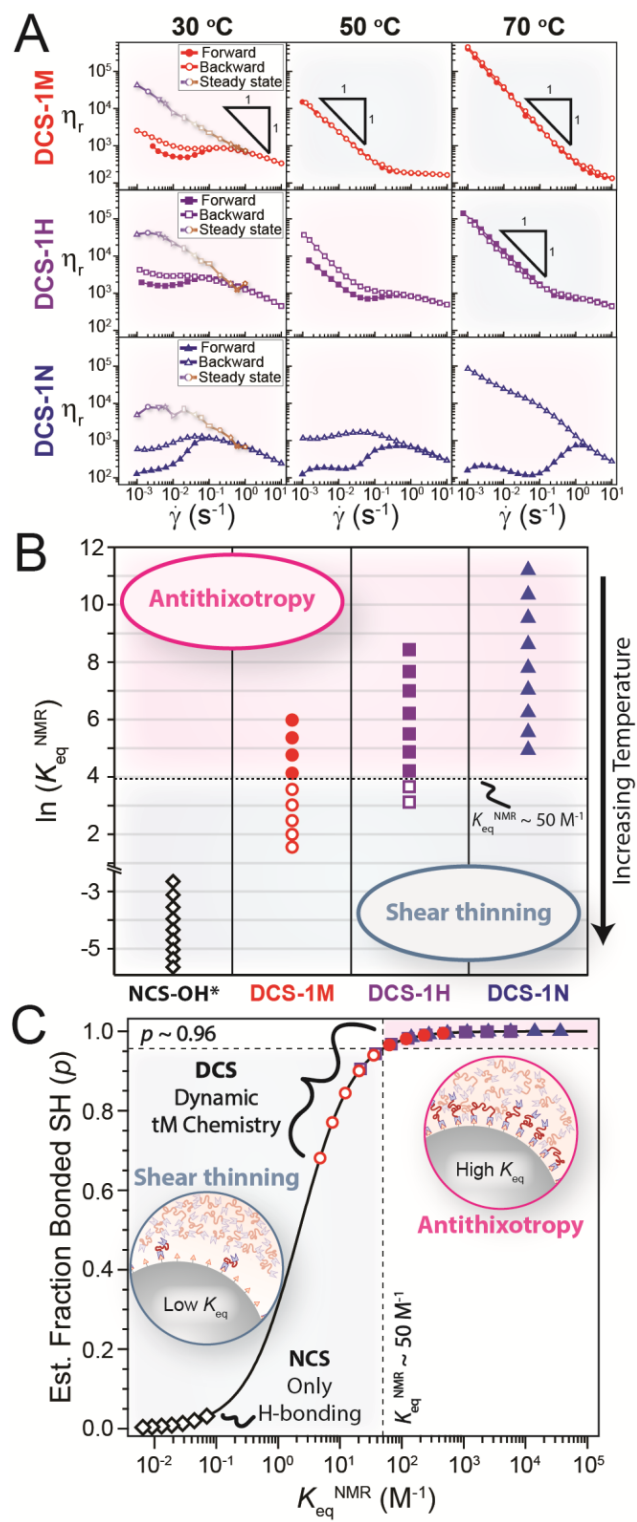

**Figure 3.** (*A*) $\eta_r$ versus $\dot{\gamma}$ for forward (*filled symbols*) and backward (*open symbols*) ramps reveal either hysteresis (antithixotropy) or no hysteresis (shear thinning). (*B*) Rheological state diagram for $0 - 80$ °C for each system showing a transition from antithixotropy (*filled symbols*) to shear thinning (*open symbols*) at $K_{eq}^{NMR} \sim 50$ M$^{-1}$ for **NCS-OH** was estimated using literature data (see Table S2). (*C*) Relating $K_{eq}^{NMR}$ to an estimated mole fraction of bound thiol ($p$), which serves as a proxy for the brush layer density (solid line, see Supporting Information for more details). Symbol identities are the same as in (*B*).

Grouping the data shown in Figure 3A in terms of $K_{eq}^{NMR}$ determined from small molecule analogs (Figure S7) yields the rheological state diagram shown in Figure 3B. Despite the potential issues in translating $K_{eq}^{NMR}$ directly to **DCSs** discussed above, the data in Figure 3B show a clear transition between antithixotropy and shear thinning at roughly the same value of $K_{eq}^{NMR}$ (~50 M$^{-1}$) whether chemistry or temperature is used as the input variable. **DCS-1N** never reaches a low enough $K_{eq}^{NMR}$ value to cross the threshold and thus only shows antithixotropy, whereas heating **DCS-1M** or **DCS-1H** leads to shear thinning. Antithixotropy in **DCSs** is accompanied with viscoelastic or liquid-like behavior under SAOS whilst shear thinning **DCSs** are solid-like (Figures S15-16). It is worth pointing out that the estimated $K_{eq}^{NMR}$ value for **NCS-OH** (Table S2) is orders of magnitude below the ~50 M$^{-1}$ threshold and this system only exhibits reversible shear thinning.

The effect of $K_{eq}$ in **DCSs** and transition from antithixotropy can be understood in terms of changes to the surface grafting density (Figure 3C), which alters particle stability in the surrounding homopolymer matrix in the quiescent, unsheared state. As detailed in the Supporting Information, $K_{eq}^{NMR}$ can be converted into the fraction of bonded thiols ($p$). Again, this $p$-value likely overestimates the bonding of polymeric tMAs to a surface due to the entropic penalty for polymer chain stretching but serves as a useful proxy for the dynamic graft density in **DCSs** (Figure 3C). Interestingly, $p$ shows a precipitous drop in the vicinity of $K_{eq}^{NMR} \sim 50$ M$^{-1}$ where **DCSs** transition from antithixotropy to shear thinning.

Such a change in nanoparticle dispersibility with grafting density is precedented for covalently grafted polymer brushes. Polymer grafted nanoparticle (PGNP) stability in a polymer melt depends on grafting density and the relative length of the grafted and matrix polymer chains.[63-65] In the special case where the graft and matrix polymer chains are the same length (as is the case here), too low of a grafting density leads to partial wetting of the brush by the matrix and too high of a grafting density causes brush dewetting or a "dry" brush. Both cases lead to particle aggregation due to an entropic depletion attraction. In contrast, intermediate grafting density leads to a wet brush and provides a repulsive barrier and particle dispersal.

Extending these lessons from covalent brushes to dynamic covalent brushes, $K_{eq}$ controls the time-averaged graft density ($p$) (Figure 3C). A large $p$ leads to an initially dispersed quiescent state as evidenced by the relatively low initial $\eta_r$ (Figure 3A) and liquid-like or viscoelastic SAOS response (Figure S15-16). Decreasing $K_{eq}$ below ~50 M$^{-1}$ leads to a precipitous drop in $p$ which drives particle-particle contacts at rest,[24, 29, 36] as evidenced by the higher $\eta_r$, shear thinning under steady shear, and solid-like SAOS response (Figures 3A and S15-16).[19, 64] The system would be expected to be in the concentrated polymer brush (CPB) regime from a minimum graft density of ~0.07 chains/nm$^2$ to the maximum theoretical graft density of 0.5 chains/nm$^2$ set by the interfacial -SH density.[66, 67]

The analogy to PGNPs applies to the initial quiescent state of the suspension (*i.e.* dispersed or aggregated), but does not answer why antithixotropy is observed for **DCSs** under steady shear. Antithixotropy requires shear-induced contacts, which could either be stabilized by polymer bridging or interparticle frictional contacts. Shear-induced polymer bridging would be possible in **DCSs** with ditopic tMAs and has been reported in "shake gels" of small ~20 nm particles with high molecular weight ~10$^6$ g/mol polymers.[68-72] On the other hand, antithixotropy is also possible without bridging interactions due to particle-particle frictional contacts, as seen in suspensions with high aspect ratio particles.[12, 73, 74] Frictional contacts would be possible in **DCSs** if the dynamic brush were to debond from the particle surface under shear and allow interparticle contacts.

To understand whether the primary mechanism for antithixotropy in **DCSs** involves polymer bridging or frictional particle-particle contacts, monotopic tMAs **3M** and **3N** (incapable of bridging) were synthesized (Figure 4A). As the molecular weight of the monotopic **3** is nominally half that of the ditopic **1**, both systems possess roughly the same stoichiometric imbalance of tMA to thiol at a constant solids volume fraction of $\phi = 55\%$. As shown in Figure 4B, **DCS-3N** exhibits shear thinning at low shear rates followed by shear thickening past ~1 s$^{-1}$. Essentially no hysteresis is observed for **DCS-3N** and the backwards flow curve almost exactly matches the steady state viscosity (Figure S17). In contrast, the same experimental conditions for **DCS-1N** lead to pronounced hysteresis over a larger shear rate range. The slight hysteresis at low shear rates observed for **DCS-3N** is also observed for a **NCS** prepared with $M_n \sim 2,000$ g/mol hydroxy-terminated PPG (Figure S17), indicating that the dynamic covalent chemistry of monotopic tMAs does not induce hysteresis. Along the same lines, **DCS-3N** and **DCS-3M** equilibrate to their steady state viscosities at orders of magnitude lower strain values (Figure S17) than **DCS-1N** (Figure S11) or **DCS-1M** (Figure 2), respectively. Therefore, the pronounced hysteresis for ditopic tMAs is primarily the result of shear-induced polymer bridging.

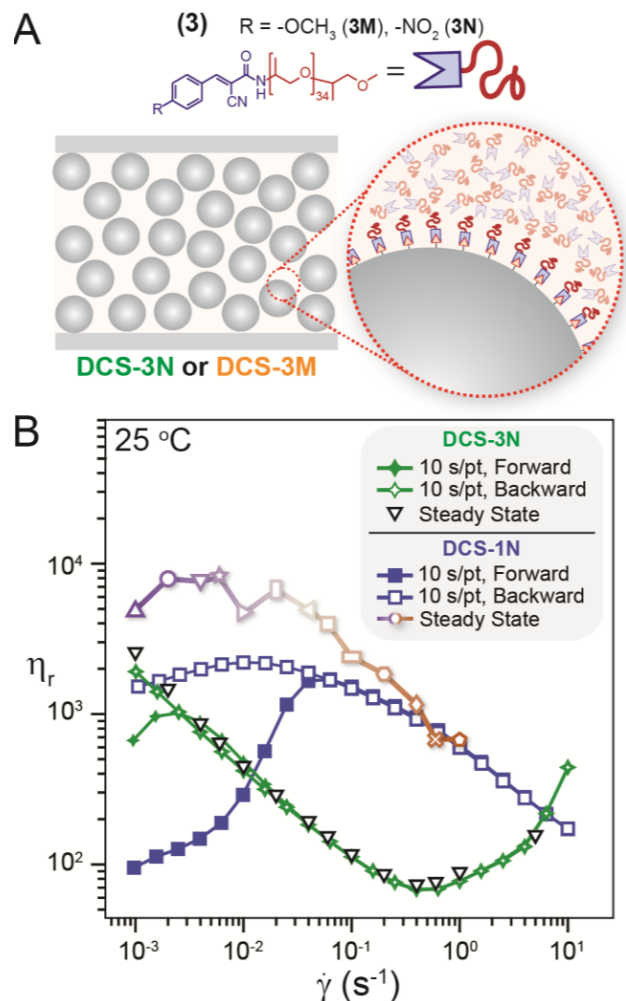

**Figure 4.** (A) A monotopic tMA **3N** or **3M** leads to **DCSs** without the possibility of shear-induced bridging between particles. (B) Reduced viscosity ($\eta_r$) for a forward-backward shear rate ($\dot{\gamma}$) ramp reveals mostly reversible behavior for **DCS-3N**, with a backward shear rate ramp which matches the steady state viscosity (Figure S17). This behavior contrasts greatly with that of the **DCS-1N**, which shows pronounced hysteresis. The steady state reduced viscosity for **DCS-3N** is also much lower than that of **DCS-1N** at a given shear rate. See Figure S17 for comparison of **DCS-3M** with **DCS-1M**.

Additionally, the ditopic **DCSs** exhibit much larger steady state viscosities than their monotopic counterparts. Even after accounting for the differences in the tMA oil viscosity, the steady state $\eta_r$ for monotopic **DCSs** at a given shear rate is approximately 20 times lower than for the ditopic **DCSs** (Figure 4B and S17). The larger steady state viscosities for ditopic **DCSs** indicates an additional attractive force between particles,[19] which again points to shear-induced polymer bridging as the main mechanism for antithixotropy.

While further experiments are needed to fully understand the effects of dynamic graft molecular weight and stoichiometric imbalance, the comparison between monotopic and ditopic tMAs demonstrates that antithixotropy in ditopic **DCSs** primarily results from shear-induced polymer bridging (Figure 5). Such a mechanism requires not only polymer grafts capable of dynamically bridging between particle surfaces, but also exposed ("bare") surface sites which are dynamically revealed by tM debonding. In this scenario, the increasing hysteresis for **DCS-1N** at elevated temperatures could be explained by a slight reduction in the dynamic graft density ($p$) which allows a larger number of polymer bridges. Additionally, Craig and coworkers have demonstrated that tensile force accelerates the dissociation rate in dynamic metal-ligand complexes,[75, 76] meaning that hydrodynamic shear stress at the particle surface could play a role in accelerating tM debonding and exposing surface sites under shear. In other words, $p$ may decrease as shear rate increases. Such tensile forces could also be responsible for the steady state shear thinning behavior of the antithixotropic networks as a large enough shear stress releases particles from their microscopic tethers before they can reform. Finally, the decay of the antithixotropic state upon shear cessation (Figure S12) correlates with the viscosity of the particle network before decay, with a lesser dependence on the viscosity of the dynamic tMA oil matrix. These data suggest that once the polymer bridges have stabilized the shear-induced network, contact relaxation requires particles to separate and subsequent reinfiltration of free tMA polymers to regenerate the repulsive brush layer (Figure 5).

## CONCLUSIONS

The use of dynamic covalent thia-Michael chemistry at the particle-solvent interface has been shown to be a new approach to control the macroscopic flow behavior of dense suspensions. Small molecule control experiments were used to understand how temperature and chemistry control the equilibrium bonding constant ($K_{eq}$). **DCSs** with a high $K_{eq}$ exhibit antithixotropy, a rare non-Newtonian behavior where viscosity increases with shearing time and relaxes upon shear cessation. Decreasing $K_{eq}$ in these **DCSs** led to more conventional rheology such as shear thinning. The changes in rheology with $K_{eq}$ are interpreted in terms of the polymer graft density at the particle surface and subsequent wetting behavior by the surrounding homopolymer matrix. Finally, monotopic tMAs were used to elucidate the primary mechanism of **DCS** antithixotropy, namely that the dynamic covalent brush layer partially debonds under shear to enable polymer bridges between particles.

Incorporation of dynamic covalent grafts at a particle surface provides a new path forward towards the general design of antithixotropic materials which can controllably adjust their dissipation over time in response to mechanical inputs. Furthermore, the tunability of this particular system enables temperature to control the polymer graft density *in situ* and further toggle between two types of non-Newtonian behaviors, antithixotropy and shear thinning. This fine level of control over energy absorption and dissipation opens the door to new materials for vibration dampening, shock absorption, and impact mitigation.

## *ASSOCIATED CONTENT*

Experimental details, synthetic schemes, suspension preparation, creep and constant shear rate measurements, decay of the antithixotropic state, temperature dependent rheology for **NCS-OH**, SAOS data and SAOS state diagram for all ditopic systems. This material is available free of charge via the Internet at http://pubs.acs.org.

## *ACKNOWLEDGMENT*

We acknowledge support from the Center for Hierarchical Materials Design (CHiMaD) under award number 70NANB19H005 (US Dept. Commerce). Additional support was provided by the Army Research Laboratory and was accomplished under Cooperative Agreement Number W911NF-20-2-0044. The views and conclusions contained in this document are those of the authors and should not be interpreted as representing the official policies, either expressed or implied, of the Army Research Laboratory, or the U.S. Government. The U.S. Government is authorized to reproduce and distribute reprints for Government purposes notwithstanding any copyright notation herein. Support from the Army Research Office under grants W911NF-19-1-0245 and W911NF-21-1-0038 and the the Division of Materials Research of the NSF (Award #2104694) are also gratefully acknowledged. This work made use of the shared facilities at the University of Chicago Materials Research Science and Engineering Center (MRSEC), supported by National Science Foundation under award number DMR-2011854. N.D.D. thanks the Pritzker School of Molecular Engineering for support through a postdoctoral fellowship. We thank Abhinendra Singh and Josh Mysona for stimulating discussions and the Huntsman Corporation for their donation of Jeffamine M-2005 as the starting material for compound **3**.

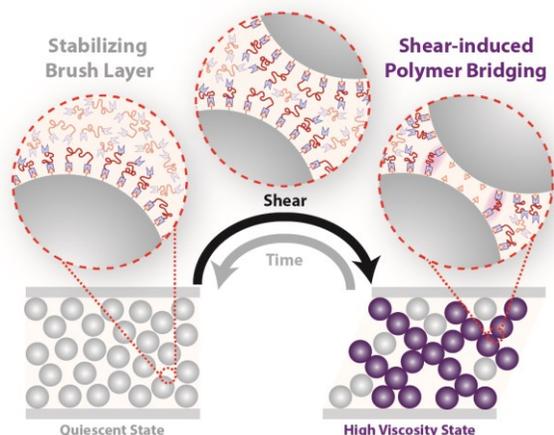

**Figure 5.** Cartoon illustration of the primary microscopic mechanism for antithixotropy in **DCSs**. Initially dispersed particles are forced into close contact by applied shear, during which the dynamic brush layer can partially debond and enable shear-induced polymer bridges which stabilize the high viscosity state. Removal of shear regenerates the sterically stabilized low viscosity quiescent state via re-bonding of free tMAs to reform a repulsive brush layer.

# Supporting Information

# Designing Stress-Adaptive Dense Suspensions using Dynamic Covalent Chemistry

Grayson L. Jackson,[a,*] Joseph M. Dennis,[b] Neil D. Dolinski,[c] Michael van der Naald,[a,d]

Hojin Kim,[a,c] Christopher Eom,[c]  Stuart J. Rowan,[c,e,f] and Heinrich M. Jaeger[a,d]

[a] *James Franck Institute, University of Chicago, 929 E. 57th Street, Chicago, Illinois 60637, United States*

[b] *Combat Capabilities and Development Command, Army Research Laboratory, Aberdeen Proving Ground, MD 21005, USA*

[c] *Pritzker School of Molecular Engineering, University of Chicago, 5640 S. Ellis Avenue, Chicago, Illinois 60637, United States*

[d] *Department of Physics, University of Chicago, 5720 S. Ellis Avenue, Chicago, Illinois 60637, United States*

[e] *Department of Chemistry, University of Chicago, 5735 S Ellis Avenue, Chicago, Illinois 60637, United States*

[f] *Chemical and Engineering Sciences Division, Argonne National Laboratory, 9700 Cass Avenue, Lemont, Illinois 60439, United States*

*Corresponding author: graysonjackson@uchicago.edu

| Table of Contents | Page |
| --- | --- |

# I.    Suspension Materials, Characterization, and Preparation

**Materials.** 500 nm silica particles (AngstromSphere Silica Microspheres) were purchased from Fiber Optic Center Inc. (New Bedford, MA, USA) and used as received. All other reagent grade solvents were purchased from Sigma-Aldrich (Milwaukee, WI, USA) and used as received.

**Characterization.** *Scanning Electron Microscopy (SEM):* SEM images were acquired using a Carl Zeiss-Merlin field emission scanning electron miscroscope housed in the Materials Preparation and Measurement Laboratory of the University of Chicago MRSEC operating at an accelerating voltage of 5.0 kV and using the Everhart-Thornley secondary electron detector.

*Nuclear Magnetic Resonance (NMR):* NMR data for dissolved particles were acquired on a 400 MHz Bruker AVANCE III HD nanobay spectrometer equipped with a BBFO SmartProbe and 24-sample SampleCase autosampler, using Topspin 3.6.2. NMR of polymer samples and variable temperature data were performed using a 500 MHz Bruker AVANCE III HD 500; 11.7 Tesla NMR at the NMR facilities at the University of Chicago. Spectra were referenced using residual protiated solvent peaks. For experimental temperatures ranging 25–40 °C, a standard plastic sample spinner was used; for equilibrium measurements at 50 °C and up, the sample was held in a ceramic spinner (to avoid spinner warpage). Sample temperatures were measured by proxy through ethylene glycol standards via procedures published previously.[1, 2]

*Size Exclusion Chromatography (SEC):* A Shimadzu HPLC operating in conjunction with a Wyatt Dawn HELEOS II multi-angle light scattering calibrated with polystyrene standards was used to determine the relative molecular weights and distributions. The eluent was THF, and the columns were an inline pair of Agilent PLgel 5 um MIXED-D columns heated to 30 °C.

*Differential Scanning Calorimetry (DSC):* DSC was performed using a TA Instruments Discovery 2500 Differential Scanning Calorimeter equipped with a refrigeration cooling unit at the Soft Matter Characterization Facility at the University of Chicago. Samples were prepared in T-zero aluminum pans from TA Instruments.  A heat-cool-heat cycle was used to eliminate thermal history of the sample and the second heat was used to determine the glass transition temperature as the midpoint of the step transition. Temperature ramps for the heat cycles were 10 °C/min, while the cooling rate was -80 °C/min.

# IA. Synthetic Methods

**Preparation of Thiol-Coated Silica Particles**

Mercaptopropyl groups were grafted onto the particle surfaces using a procedure adapted from Crucho *et al.*[3] In a 2 L flask fitted with a reflux condenser, Silica particles (15.9 g) were dispersed via sonication in 800 mL of toluene that had been dried over $MgSO_4(s)$. To this 240 mL of (3-mercaptopropyl)trimethoxysilane (10-fold stoichiometric excess) was added and heated to reflux for 24 hr under a $N_2$ atmosphere. The thiol-functionalized silica NPs were isolated via centrifugation and then redispersed in EtOH. After redispersing in EtOH and centrifuging 5 times, the supernatant tested negative for free thiols using 5,5'-dithiobis(2-nitrobenzoic acid) (*Ellman's reagent*). Particle density was determined to be 1.92 +/- 0.01 g/mL by dispersing the particles in a series of bromoform:methanol mixtures and centrifuging to see whether the particles floated or sank.

**Scheme S1.** Synthesis of thiol-coated silica particles

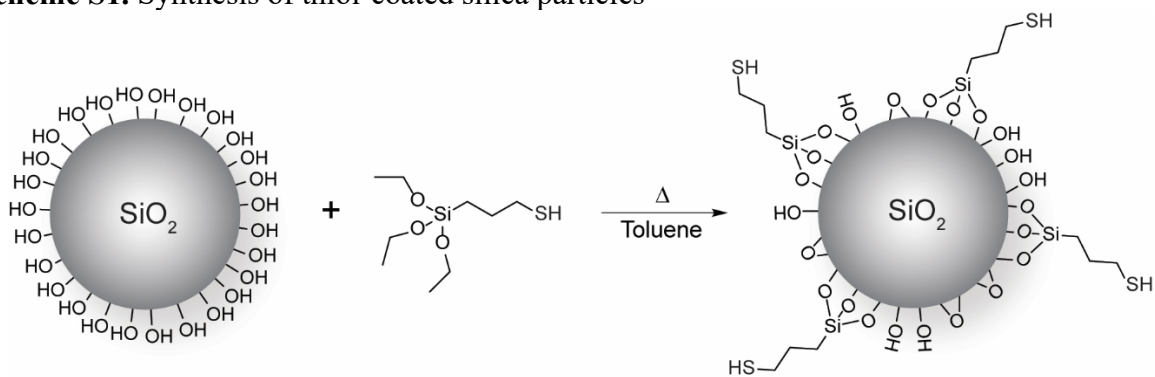

The thiol surface functional group density was determined by dissolving 25 mg of particles in 0.75 mL of 0.5 M NaOD in $D_2O$ at 90 °C and using 1,3,5-trixoane as an internal standard to measure the intensity of the HSC*H*$_2$CH$_2$CH$_2$Si(ONa)$_3$ peak at 2.82 ppm,[3] which yielded a value of 6.7 μmol SH/g particle. Using the measured particle density and assuming an idealized spherical particle, the surface density of thiol groups was calculated as 0.5 groups/nm$^2$ (~0.9 μmol/m$^2$) which is the same as that reported by Crucho *et al.*[3] and comparable to other reported values.[4] Considering the tridentate binding of the mercaptopropyl groups and a hydroxl surface group density of 5/nm$^2$,[5] our calculated value indicates that the thiol surface coverage is ~40% of its theoretical maximum as depicted schematically in Scheme S1. Note that the much smaller surface area to volume ratio for these relatively large, 417 nm diameter particles meant that surface group quantitation by FTIR[4] did not yield a discernable signal and mass loss from TGA of thiol-coated particles was indistinguishable from that due to silica network condensation in unfunctionalized particles.[3]

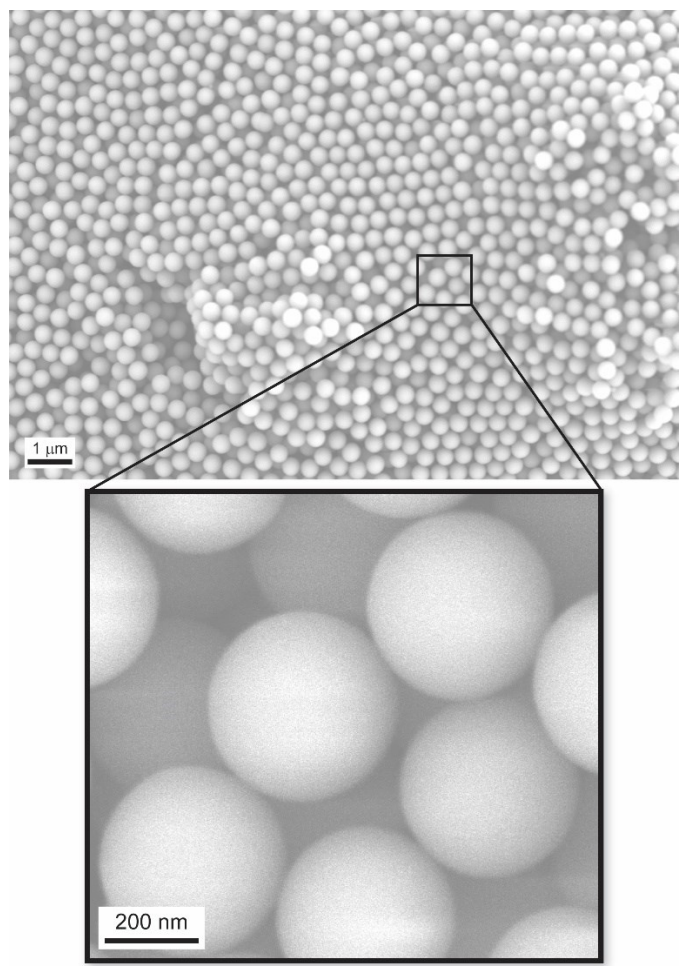

**Figure S1.** SEM of thiol-coated particles reveal nominally smooth particles with $d = 417 \pm 30$ nm.

**Synthesis of Ditopic (1M, 1H, 1N) or Monotopic (3M, 3N) Polymer tMAs for Rheology**

**Scheme S2.** Synthesis of Cyanoacetamide-terminated Polypropylene Glycol

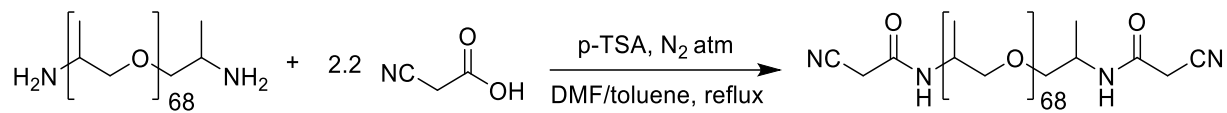

Amine-terminated poly(propylene glycol), (Jeffamine D-4000), was used as received from Sigma-Aldrich (starting material for **1M, 1H, 1N**) Monofunctional Jeffamine M-2005 was supplied by the Huntsman Corporation (starting material for **3M, 3N**). The synthetic protocols were similar for both starting materials so the functionalization of Jeffamine D-4000 is used as a representative example. Jeffamine D-4000 (10 g, 2.5 mmol), *para*-toluene sulfonic acid (*p*-TSA) (0.1 g) and cyanoacetic acid (0.47 g, 5.5 mmol) was added to a 100-mL, round-bottomed, two-necked flask equipped with a magnetic stir bar, nitrogen inlet, temperature probe and Dean-stark trap with condenser. Dimethylformamide (DMF) (10 mL) and toluene (30 mL) were added to the flask and the contents were purged with nitrogen for 10 minutes at room temperature. Following the nitrogen purge, the homogeneous solution was then heated to a reflux temperature of 110 °C, as measured by the temperature probe. The reaction proceeded under reflux until the azeotropic removal of water into the Dean-stark trap ceased (approximately 4 hours). In general, the quantitative measure of water removal corresponded well to the expected reaction conversion with an additional amount of water likely resulting from water contamination of the starting materials (e.g. DMF and PPG). After 4 hours, toluene was then distilled from the reaction flask until the solution temperature reached 120 °C (approximately 30 min). The resulting homogenous, pale yellow solution was then cooled to room temperature and diluted with 50 mL of chloroform. Three washes of a saturated NaHCO$_3$(*aq*) solution, followed by three washes with deionized water (100 mL each wash) were used to remove the residual *p*-TSA and DMF. Finally, the organic phase was then dried with MgSO$_4$(*s*), filtered, and the remaining chloroform was removed using a rotary evaporator.

**Yield:** 9.2 g (92%) of a pale yellow/orange viscous oil

**Scheme S3.** Synthesis of Ditopic (**1M**, **1H**, **1N**) or Monotopic (**3M, 3N**) tMA (Benzalcyanoacetamide-terminated Polypropylene Glycol)

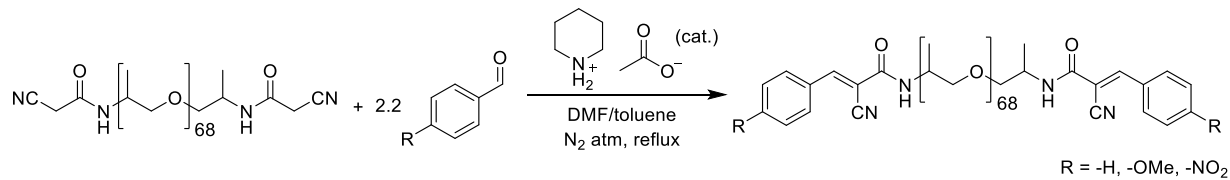

R = -H, -OMe, -NO$_2$

Conversion of the cyanoacetamide-terminated PPG to a ditopic (**1M, 1H, 1N**) or monotopic (**3M, 3N**) benzalcyanoacetamide-terminated poly(propylene glycol) followed a similar procedure described by Cope.[6] As a representative example, the synthesis of the unsubstituted benzalcyanoacetamide-terminated PPG is described. Cyanoacetamide-terminated poly(propylene glycol) (10 g, 2.5 mmol), benzaldehyde (0.58 g, 5.5 mmol) and piperdinium acetate (0.1 g) were added to a 100-mL, round-bottomed, two-necked flask equipped with a magnetic stir bar, nitrogen inlet, temperature probe, and Dean-stark trap with condenser. DMF (10 mL) and toluene (30 mL) were added to the flask and the contents were purged with nitrogen for 10 minutes at room temperature. Following the nitrogen purge, the homogeneous solution was then heated to a reflux temperature of 110 °C, as measured by the temperature probe. The reaction proceeded under reflux until the azeotropic removal of water into the Dean-stark trap ceased (approximately 4 hours). After 4 hours, toluene was then distilled from the reaction flask until the solution temperature reached 120 °C (approximately 30 min). The resulting homogenous, pale orange solution was then cooled to room temperature and diluted with 50 mL of chloroform. Three washes of a saturated sodium bicarbonate solution, followed by three washes with deionized water (100 mL each wash) removed the residual benzaldehyde and DMF. Finally, the organic phase was then dried with $MgSO_4(s)$, filtered, and the remaining chloroform was removed using a rotary evaporator. Products were recovered as a pale orange/brown viscous oils.

**1M**: Recovered Yield: 82% [1]H NMR (500 MHz, CDCl$_3$): δ = 8.26 (s, 1H, C=CH), 7.72 (d, 2H, Ar-H), 6.93 (d, 2H, Ar-H), 3.87 (s, 3H, Ar-OCH$_3$), 3.36-3.66 (m, -CH$_2$-CH-), 1.26 (m, -CH$_3$) ppm.

**1H**: Recovered Yield: 86% [1]H NMR (500 MHz, CDCl$_3$): δ = 8.34 (s, 1H, C=CH), 7.75 (t, 3H, Ar-H), 7.48 (d, 2H, Ar-H), 3.36-3.66 (m, -CH$_2$-CH-), 1.26 (m, -CH$_3$) ppm.

**1N**: Recovered Yield: 75% [1]H NMR (500 MHz, CDCl$_3$): δ = 8.46 (s, 1H, C=CH), 8.26 (d, 2H, Ar-H), 7.91 (d, 2H, Ar-H), 3.36-3.66 (m, -CH$_2$-CH-), 1.26 (m, -CH$_3$) ppm.

**3M**: Recovered Yield: 85% [1]H NMR (500 MHz, CDCl$_3$): δ = 8.26 (s, 1H, C=CH), 7.68 (d, 2H, Ar-H), 6.91 (d, 2H, Ar-H), 3.86 (s, 3H, Ar-OCH$_3$), 3.36-3.66 (m, -CH$_2$-CH-), 1.26 (m, -CH$_3$) ppm.

**3N**: Recovered Yield: 71% [1]H NMR (500 MHz, CDCl$_3$): δ = 8.39 (s, 1H, C=CH), 8.26 (d, 2H, Ar-H), 7.90 (d, 2H, Ar-H), 3.36-3.66 (m, -CH$_2$-CH-), 1.15 (m, -CH$_3$) ppm.

**Table S1.** Polymer Characterization

| | GPC | | NMR | T$_g$ |
|---|---|---|---|---|
| Sample | M$_n$ (x $10^3$ g/mol) | Đ | MW (x $10^3$ g/mol) | °C |
| **1M** | 6.3 | 1.51 | 6.5 | -67 |
| **1H** | 6.2 | 1.52 | 6.3 | -67 |
| **1N** | 5.7 | 1.48 | 7.0 | -65 |
| **3M** | 3.6 | 1.16 | 3.4 | -70 |
| **3N** | 4.0 | 1.10 | 2.7 | -72 |

**Synthesis of Small Molecule tMAs (2) for NMR Experiments:**

**Scheme S4. Synthesis of 2-cyano-N-ethylacetamide precursor:**

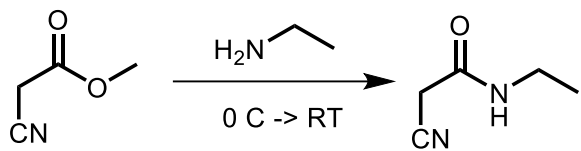

Synthesis adapted from Ref[7]. Methyl cyanoacetate (9.9 g, 99.9 mmol) was added to a 250 mL flask and 15 mL of ethyl amine solution (68-72% in water) was added dropwise (~15 minutes to completely add) at 0 C. After addition, the ice bath was removed and the reaction mixture was allowed to stir for 12 hours. The mixture was then evaporated under reduced pressure to yield a thick, dark brown oil. At this point, 20 mL of ether and 10 mL of hexanes were added, precipitating an orange solid. The mixture was then filtered and washed with 2:1 ether:hexanes to yield the desired solid product, which was used without further purification.

**Yield:** 10 g (89%) of an orange solid
**[1]H NMR** (500 MHz, CDCl$_3$): δ 7.20 (s, 1H, -NH-), 3.44 (s, 2H, -CH$_2$-), 3.30 – 3.20 (m, 2H, -CH$_2$-), 1.12 (t, $J$ = 7.3 Hz, 3H, -CH$_3$) ppm.
**[13]C NMR** (126 MHz, CDCl$_3$): δ 161.83, 115.20, 35.22, 25.95, 14.28 ppm.

**Scheme S5. Synthesis of 2R from 2-cyano-N-ethylacetamide**

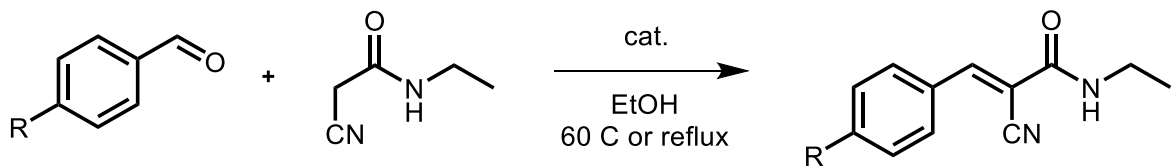

*Synthesis of 2M:*

2-cyano-N-ethylacetamide (0.5 mg, 4.46 mmol) and p-anisaldehyde (542 µL, 4.46 mmol) were dissolved in 20 mL of ethanol. The reaction was heated to reflux, at which point ~20 mg of piperidinium acetate was added to the solution, and was left to react for 12 hours. The reaction was then cooled, and the solvent was evaporated. The residue was redissolved in methanol and placed in a freezer (-20 C) for several hours. Water was added to the cooled solution (~10-20% volume) until a precipitate formed. The mixture was then filtered and washed with 1:1 water:methanol to yield the desired solid product.

**Yield:** 350 mg (33%) of an off-white solid
**[1]H NMR** (500 MHz, CDCl$_3$): δ 8.25 (s, 1H, C=CH), 7.93 (d, $J$ = 8.9 Hz, 2H, Ar-H), 6.98 (d, $J$ = 8.9 Hz, 2H, Ar-H), 6.33 (s, 1H, -NH-), 3.88 (s, 3H, -CH$_3$), 3.47 (qd, $J$ = 7.3, 5.6 Hz, 2H, -CH$_2$-), 1.24 (t, $J$ = 7.3 Hz, 3H, -CH$_3$) ppm.
**[13]C NMR** (126 MHz, CDCl$_3$): δ 163.35, 160.75, 152.27, 133.12, 124.80, 117.88, 114.77, 100.68, 55.66, 35.54, 14.80, 14.79 ppm.

### Synthesis of 2H:

2-cyano-*N*-ethylacetamide (1.0515 g, 9.3775 mmol) and benzaldehyde (953.2 μL, 9.3775 mmol) were combined with piperidine (926.33 μL, 9.3775 mmol) in 282.66 mL ethanol. The reaction mixture was stirred for 12 h at 60 C, and the solvent was evaporated. The residue was purified by column chromatography on silica gel ($CH_2Cl_2$).

**Yield:** 710.6 mg (38%) of a bright yellow solid
**¹H NMR** (500 MHz, $CDCl_3$): δ 8.34 (s, 1H, C=CH), 7.93 (d, *J* = 6.8 Hz, 2H, Ar-H), 7.57 – 7.45 (m, 3H, Ar-H), 6.41 (s, 1H, -NH-), 3.53 – 3.44 (m, 2H, -CH$_2$-), 1.26 (t, *J* = 7.3 Hz, 3H, -CH$_3$).
**¹³C NMR** (126 MHz, $CDCl_3$) δ 160.10, 152.94, 132.79, 131.97, 130.68, 129.32, 117.20, 104.21, 35.64, 14.74.

### Synthesis of 2N:

2-cyano-N-ethylacetamide (0.5 mg, 4.46 mmol) and 4-nitrobenzaldehyde (674 mg, 4.46 mmol) were dissolved in 20 mL of ethanol. The reaction was heated to reflux, at which point ~20 mg of piperidinium acetate was added to the solution and was left to react for 12 hours. The reaction was then cooled, and the solvent was evaporated. The residue was redissolved in methanol and placed in a freezer (-20 C) for several hours. Water was added to the cooled solution (~10-20% volume) until a precipitate formed. The mixture was then filtered and washed with 1:1 water:methanol to yield the desired solid product.

**Yield:** 450 mg (41%) of a pale-yellow solid
**¹H NMR** (500 MHz, $CDCl_3$): δ 8.39 (s, 1H, C=CH), 8.33 (d, *J* = 8.8 Hz, 2H, Ar-H), 8.06 (d, *J* = 8.8 Hz, 2H, Ar-H), 6.47 (d, *J* = 5.8 Hz, 1H, -NH-), 3.50 (qd, *J* = 7.3, 5.6 Hz, 2H, -CH$_2$-), 1.27 (t, *J* = 7.3 Hz, 3H, -CH$_3$) ppm.
**¹³C NMR** (126 MHz, $CDCl_3$): δ 158.94, 149.81, 149.52, 137.57, 131.17, 124.41, 116.18, 108.53, 35.87, 14.65 ppm.

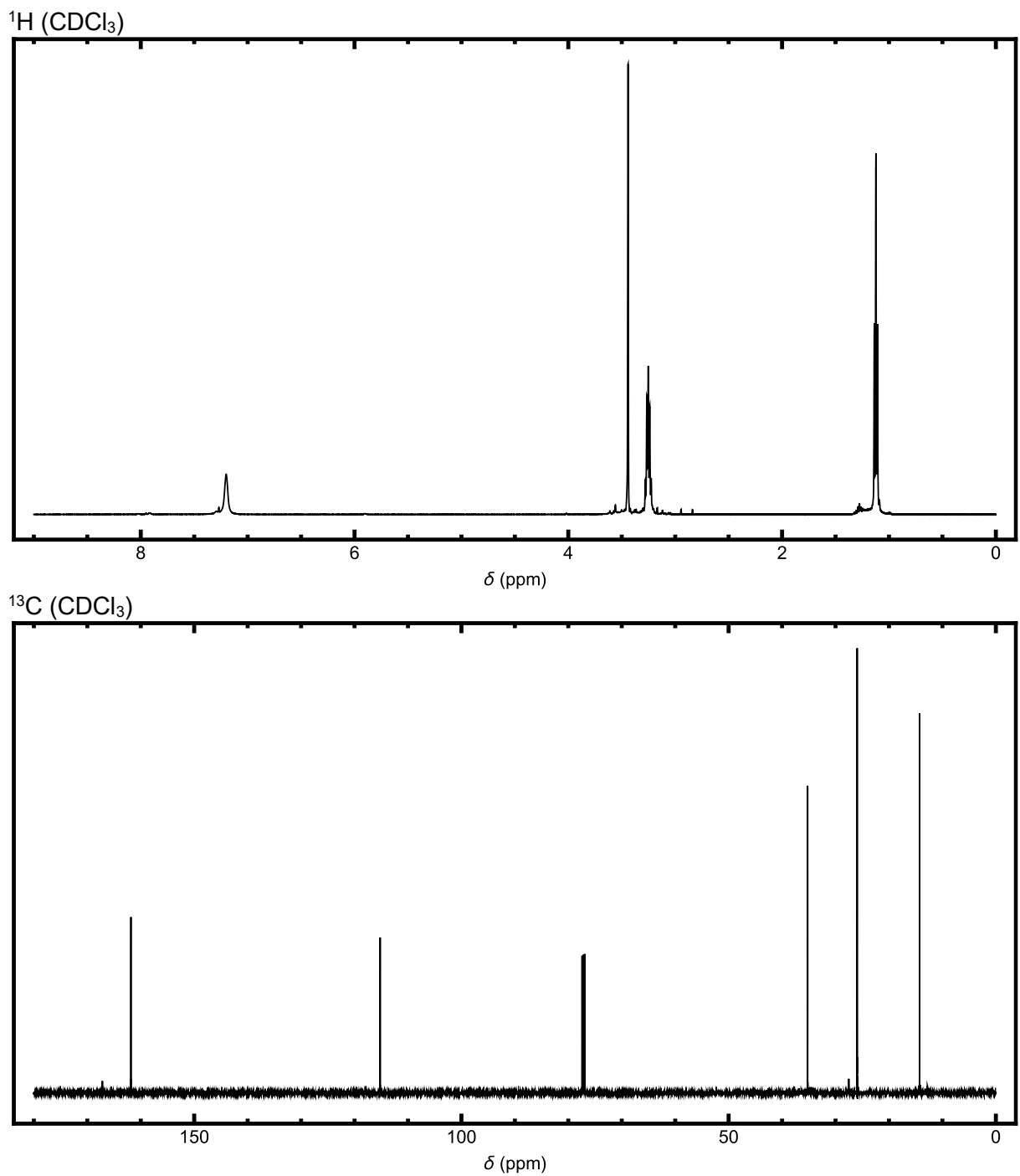

**Figure S2.** [1]H and [13]C NMR spectra for 2-cyano-N-ethylacetamide

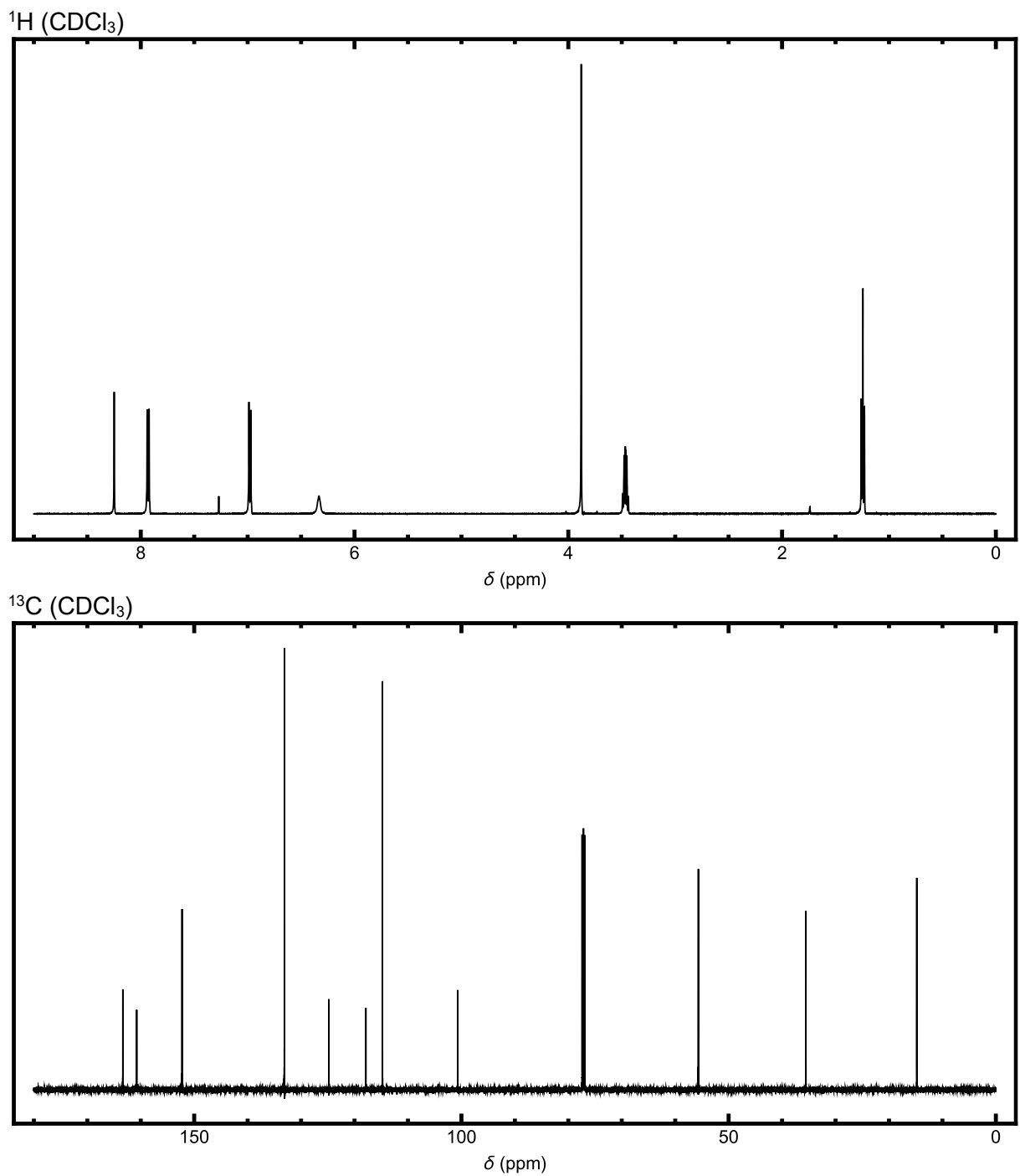

**Figure S3.** [1]H and [13]C NMR spectra for **2M**.

<sup>1</sup>H (CDCl<sub>3</sub>)

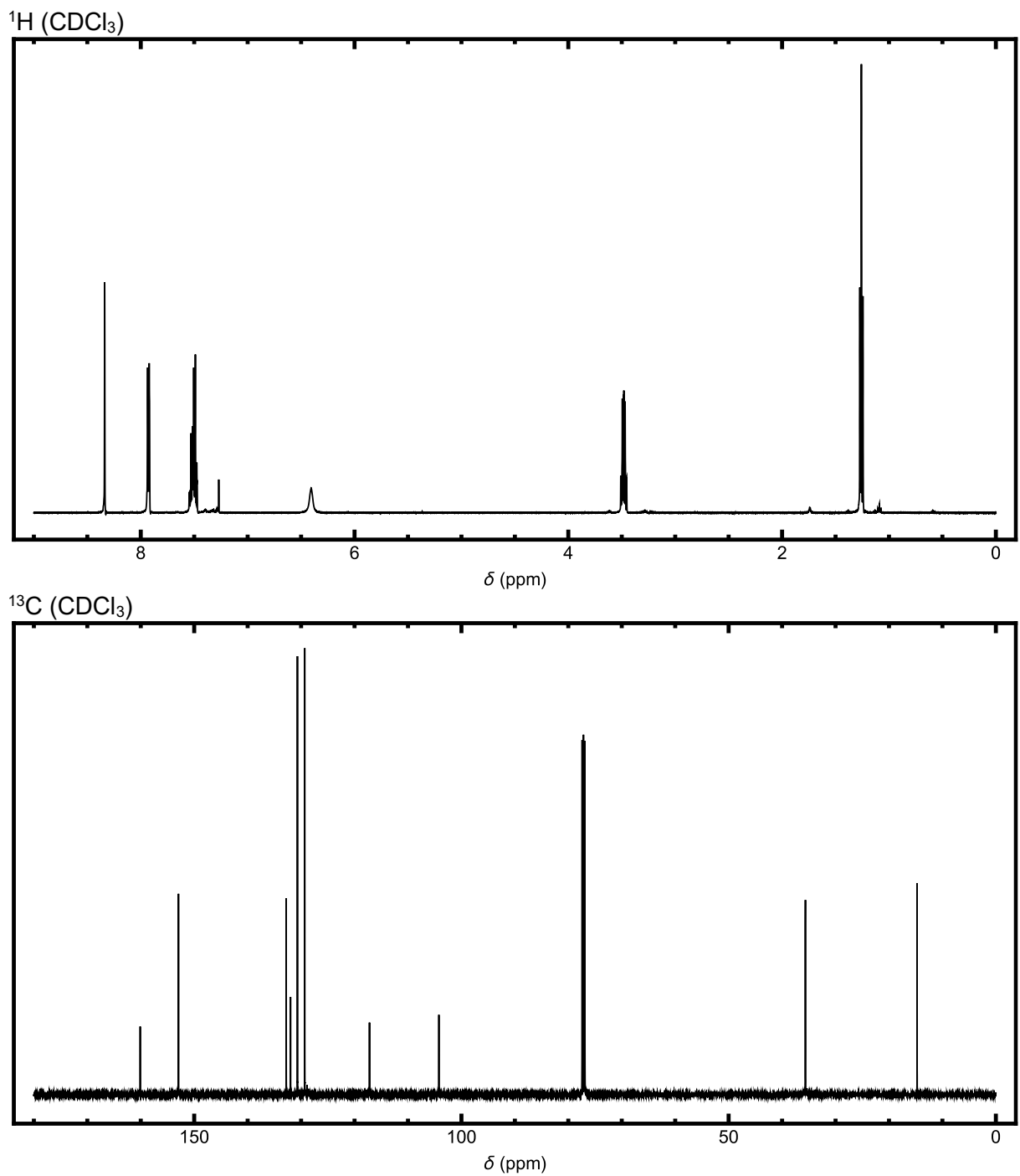

<sup>13</sup>C (CDCl<sub>3</sub>)

**Figure S4.** <sup>1</sup>H and <sup>13</sup>C NMR spectra for **2H**.

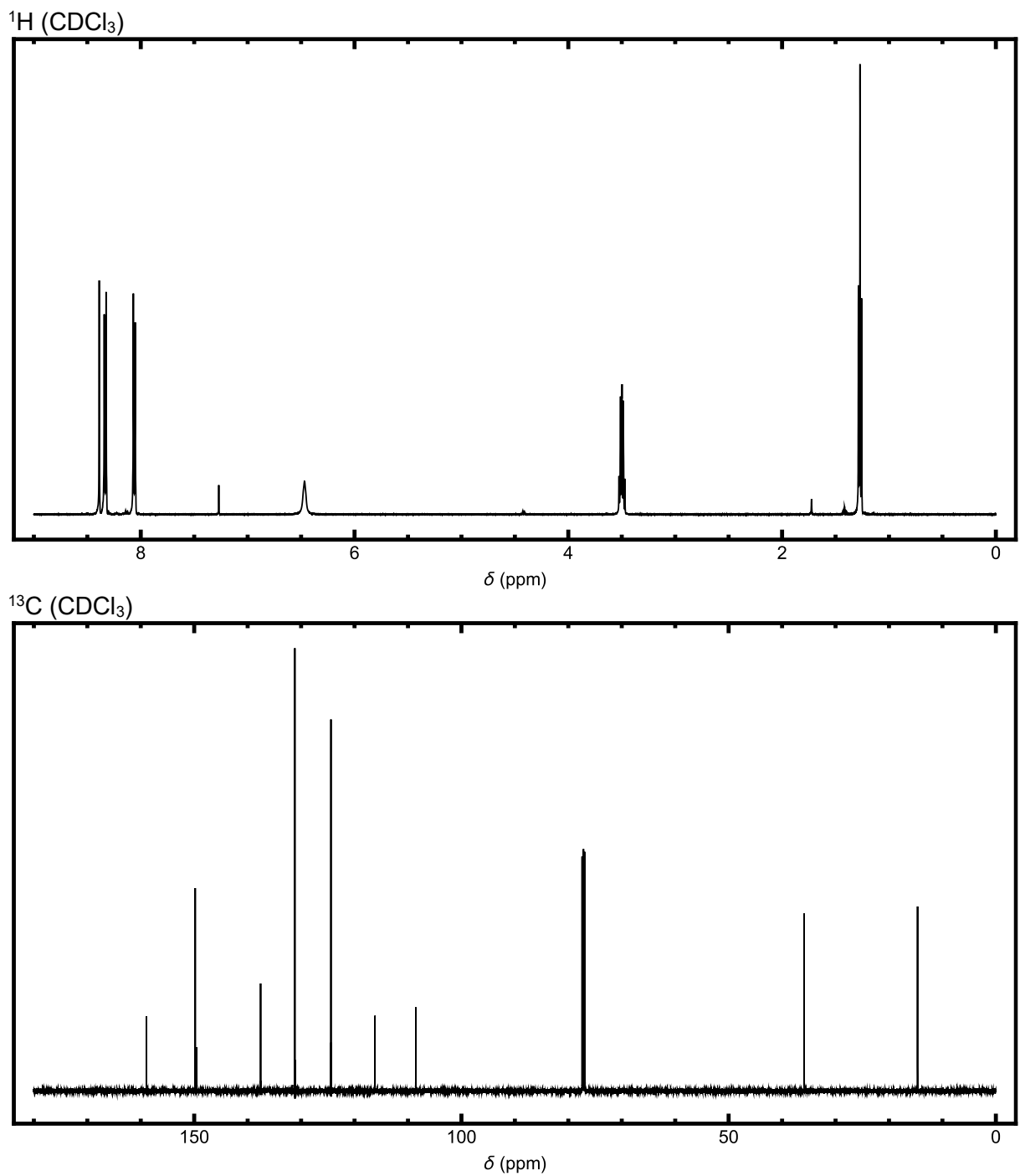

**Figure S5.** [1]H and [13]C NMR spectra for **2N**.

## IB. Determination of $K_{eq}$ from small molecule analogs (2M, 2H, 2N)

Stock solutions (200 mM, ~0.5 mL) of the benzalcyanoacetamide species of interest (**2M, 2H, 2N**) and 1-octanethiol were prepared in $d_6$-DMSO from freshly opened ampules in an inert (glovebox) environment. Equal volumes of each solution (255 µL) were then added to an intermediary vial and were vigorously mixed. Afterwards, the mixture was added to an NMR tube via a syringe and allowed to equilibrate for 12 hours. The sample was then introduced into the (preheated) NMR and allowed to equilibrate to the environmental temperature for 5 minutes before collecting data to ensure homogeneous sample temperature. Samples were monitored until equilibrium was reached (minimum equilibration time of 30 minutes). Afterwards, the temperature of the NMR would be increased, and the equilibration procedure would be repeated for all desired temperatures. Representative peaks of interest are displayed below in Figure S6, and the governing equation(s) used for monitoring adduct formation are also reported below.

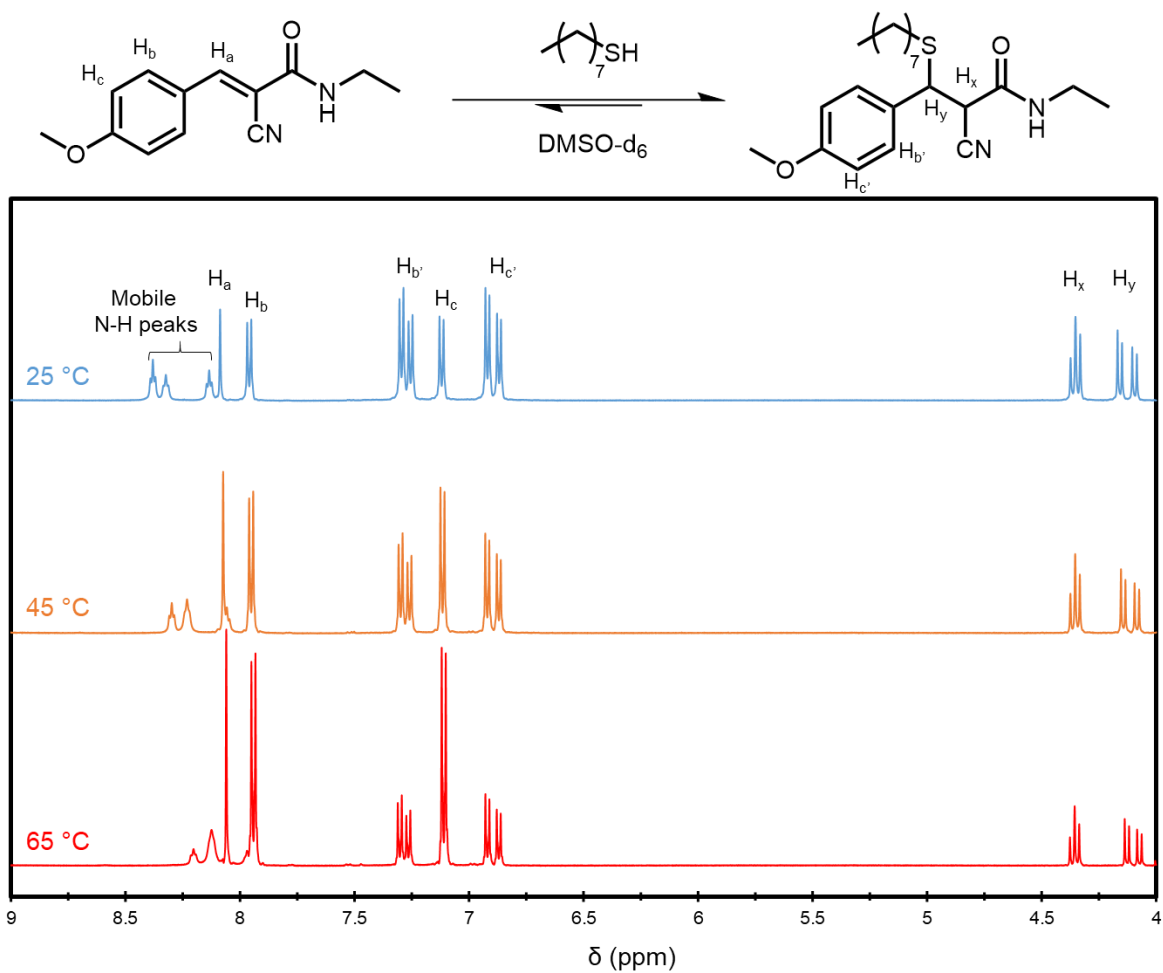

$$mol\ frac\ bound = \frac{\int H_x + \int H_y}{\int H_x + \int H_y + 2\int H_a} = \frac{\int H_x + \int H_y}{\int H_x + \int H_y + \int H_b} = \frac{\int H_x + \int H_y}{\int H_x + \int H_y + \int H_c}$$

$$K_{eq} = \frac{[bound\ species]}{[acceptor][thiol]} = \frac{sample\ concentration * (mol\ frac\ bound)}{(sample\ concentration * (1 - mol\ frac\ bound))^2}$$

**Figure S6.** Representative equilibrium NMR spectra for **2M** and 1-octanethiol obtained during at several temperatures (100 mM in DMSO-d$_6$) along with formulas to obtain the mole fraction (*mol frac*) of bonded species and $K_{eq}$. It is important to note that increasing temperature shifts the position of the N-H peaks, which will convolute integrated values, requiring the use of several peaks.

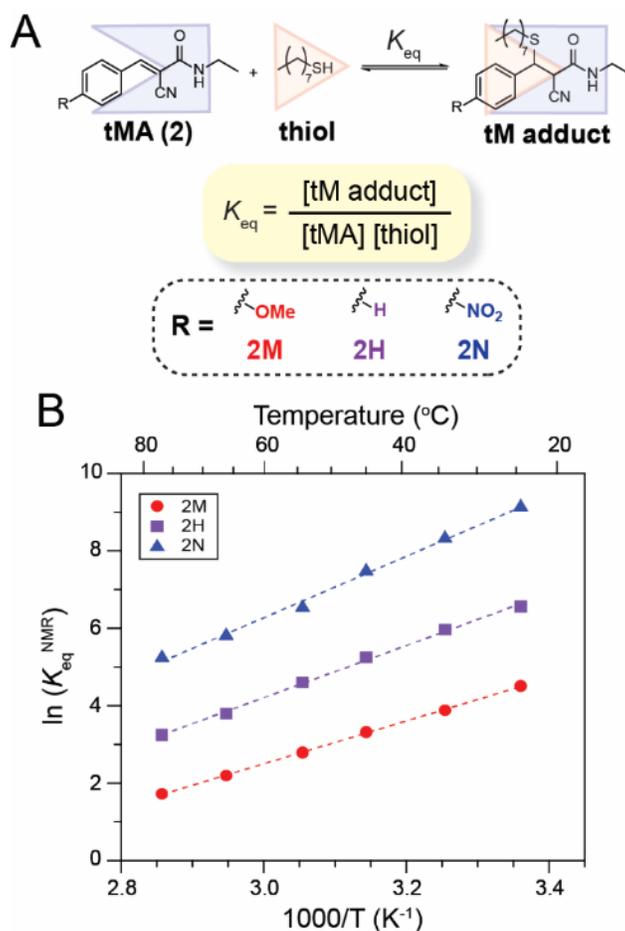

**Figure S7.** Small molecule analogs (**2M**, **2H**, **2N**) were used to determine the equilibrium bonding constant ($K_{eq}^{NMR}$) as a function of tMA chemistry and temperature. (*A*) Schematic of benzalcyanoacetamide electrophile (*tMA*) equilibrium reaction with 1-octanethiol (*thiol*) to form a thioether (*tM adduct*). (*B*) $K_{eq}^{NMR}$ as a function of temperature and R-substituent at 100 mM equimolar conditions in DMSO. The dashed line is an Arrhenius fit to the data and the van't Hoff fit parameters are displayed in Table S2.

**Table S2.** Results of linear fits to the van't Hoff data (error reported from standard error of fit). See Herbert *et al*.[1] for additional details of $K_{eq}$ measurement by NMR.

| Compound | ΔH (kJ mol⁻¹) | ΔS (J mol⁻¹ K⁻¹) |
|---|---|---|
| **2M** | -46.0 ± 0.4 | -117 ± 1 |
| **2H** | -56.0 ± 1.3 | -133 ± 4 |
| **2N** | -65.9 ± 2.1 | -145 ± 7 |
| **OH** | -33* | -132** |

*Due to the presence of thiol groups as well as likely residual silanol groups on the particle surface, the ΔH value was estimated using the average hydrogen bond strength for O-H•••O and O-H•••S of alcohols and thiols as H-bond acceptors.[8]
**This ΔS value was estimated as the average of the **2M**, **2H**, and **2N** compounds.

## II. Suspension Preparation

Thiol-functionalized particles and tMAs (**1M**, **1H**, **1N**, **3N**, or **3M**) or 4,000 g/mol hydroxy-terminated PPG (**OH**) were massed into 4 dram vials and homogenized by iterative hand mixing, centrifugation, and water bath sonication at 40 kHz and 130 W. Once homogeneous, the samples were annealed at 40 °C in the sonicator for 12 h and then allowed to rest for at least one week at ambient temperature prior to rheological characterization. Volume fractions were calculated using either $\rho_{2OH4000} = 1.004$ g/mL or $\rho_{tMA} = 1$ g/mL and the experimentally determined value of $\rho_{Particles} = 1.92$ g/mL.

**Estimation of Fraction Bonded SH ($p$) for DCSs from $K_{eq}^{NMR}$:**

The equation for the equilibrium association constant ($K_{eq}^{NMR}$) is given by Equation 1:

$$K_{eq}^{NMR} = \frac{[tM\ adduct]}{[free\ SH][free\ tMA]} \tag{1}$$

As $K_{eq}^{NMR}$ was determined under equimolar conditions, it cannot directly be related to the mole fraction of bound thiol ($p$) in **DCSs** (which are at stoichiometric imbalance where the thiol (SH) is the limiting reagent) as indicated by the equation in Figure S6. Instead, the concentrations of the various components are given by Equations 2-4:

$$[tM\ adduct] = [SH]_0 * p \tag{2}$$

$$[free\ SH] = [SH]_0 * (1 - p) \tag{3}$$

$$[free\ tMA] = [tMA]_0 - [tM\ adduct] = [tMA]_0 - [SH]_0 * p \tag{4}$$

Where $[SH]_0$ and $[tMA]_0$ correspond to the initial concentrations of 0.016 and 0.46 M in **DCSs**, respectively. Substituting Equations 2-4 into Equation 1 yields Equation 6:

$$K_{eq}^{NMR} = \frac{p}{(1-p)*([tMA]_0 - [SH]_0 * p)} \tag{6}$$

Using the $K_{eq}^{NMR}$ values, this equation was solved analytically to yield $p$ and is plotted as the solid curve in Figure 3C in the main text.

# III. Rheological Characterization

All rheological measurements were conducted using either an Anton Paar MCR301 or MCR702 rheometer with a smooth parallel plate geometry (d = 25 mm) using a gap size of approximately 0.1 mm. Temperature was controlled with a precision of 0.1 °C using a Peltier device and water circulator. The apparent viscosity of the suspensions ($\eta$) was divided by that of the background solvent ($\eta_0$, Figure S8) to attain the reduced viscosity ($\eta_r$) and isolate the viscosity contribution of the particles.

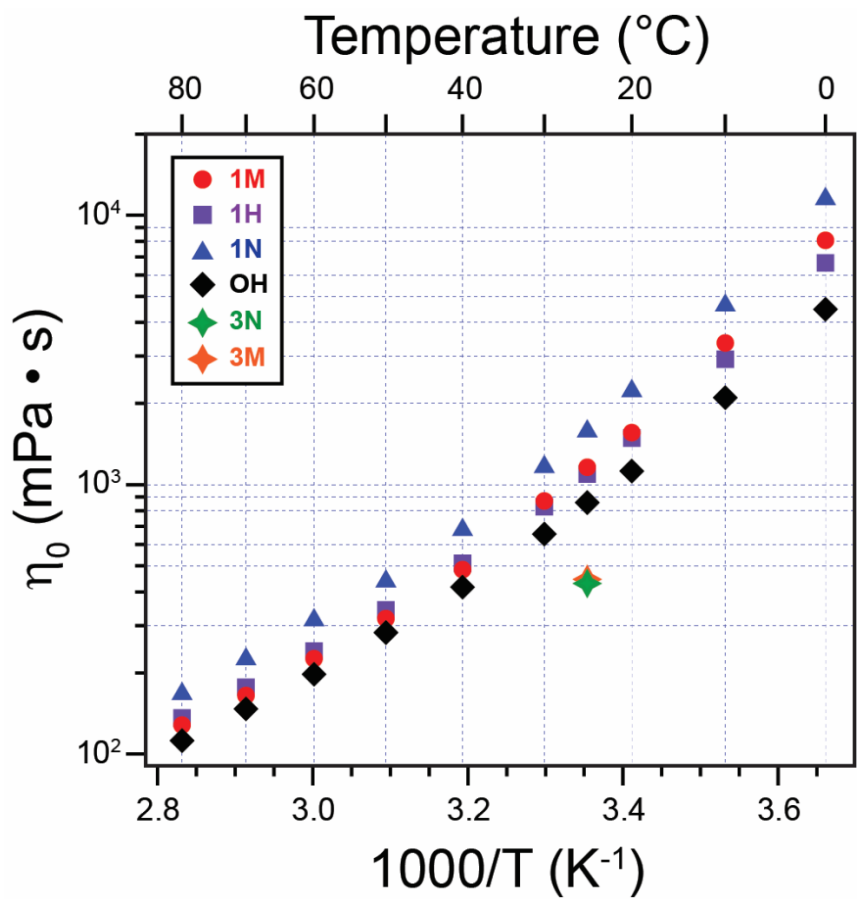

**Figure S8.** Viscosity ($\eta_0$) of ditopic tMAs (**1M**, **1H**, **1N**), monotopic tMA (**3M**, **3N**), and hydroxy-terminated 4000 g/mol polypropylene glycol (**OH**) as a function of temperature. The viscosities were Newtonian over the measured shear rate range of 0.1 – 100 s$^{-1}$.

For the hysteresis loop experiments, a preshear of either 20 s$^{-1}$ (data in Figure 2) or 5 s$^{-1}$ (data in Figures 3 & 4) were applied for 120 s, followed by a rest period of 5 minutes and then the shear rate sweep was conducted.

For the temperature-dependent measurements, the samples were equilibrated at each temperature for 600 s with $\omega$ = 10 Hz and $\gamma$ = 0.5%. The oscillatory measurements were conducted from $\omega$ = 100 to 0.1 Hz with $\gamma$ = 0.5%, which took approximately 1800 s, and then the hysteresis loop measurements were performed in triplicate with a waiting time of 10 s/pt. Temperature dependent rheology exhibited qualitatively the same rheological behavior on cooling.

For the constant shear rate or constant shear stress (*creep*) measurements shown in Figure 2 and Figures S9-11, a preshear of 20 s$^{-1}$ was applied for 120 s and then the shear rate or stress was immediately reduced to the indicated value. The highest particle Reynold's number reached in our experiments, which occurred at the highest temperatures where the dynamic oil viscosity was the lowest, was 10$^{-8}$ <<<<< 1, meaning that particle inertia can be safely neglected.

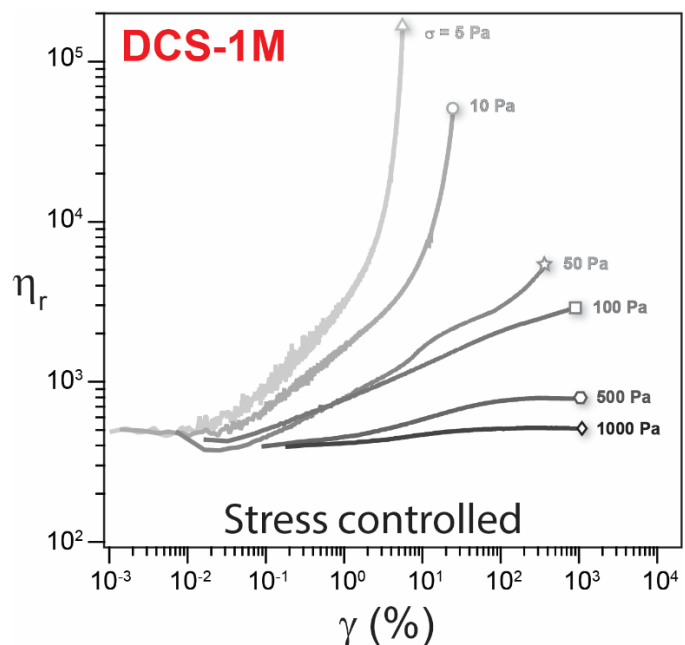

**Figure S9.** Constant stress (*creep*) measurements of **DCS-1M** at 25 °C showing the change in the reduced viscosity ($\eta_r$) as a function of strain ($\gamma$). The applied preshear was 20 s$^{-1}$ for 120 s and then the shear stress ($\sigma$) was immediately set to the indicated value. The system exhibits a viscosity bifurcation,[9] wherein the viscosity diverges for $\sigma \leq 10$ Pa and flows for $\sigma \geq 100$ Pa.

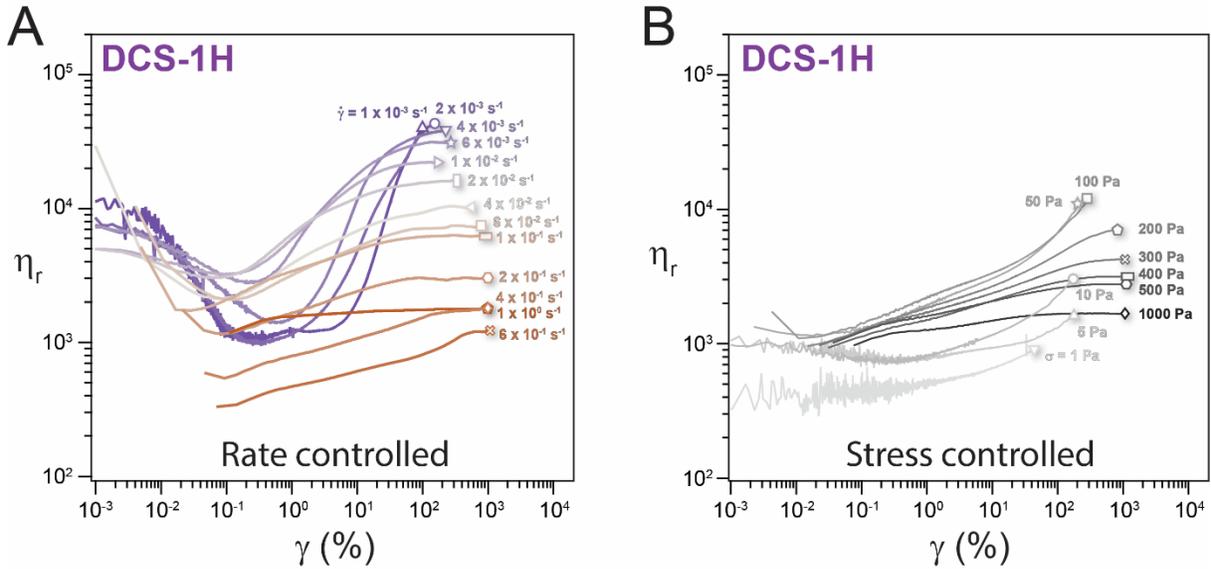

**Figure S10.** (*A*) Constant shear rate and measurements of **DCS-1H** at 25 °C. Reduced viscosity (η_r) increases with strain (γ) and at most rates plateaus to a high γ viscosity value. These high γ values are plotted in Figure 3A as the 25 °C steady state for **DCS-1H**. (*B*) Constant shear stress (*creep*) measurements show a viscosity bifurcation.

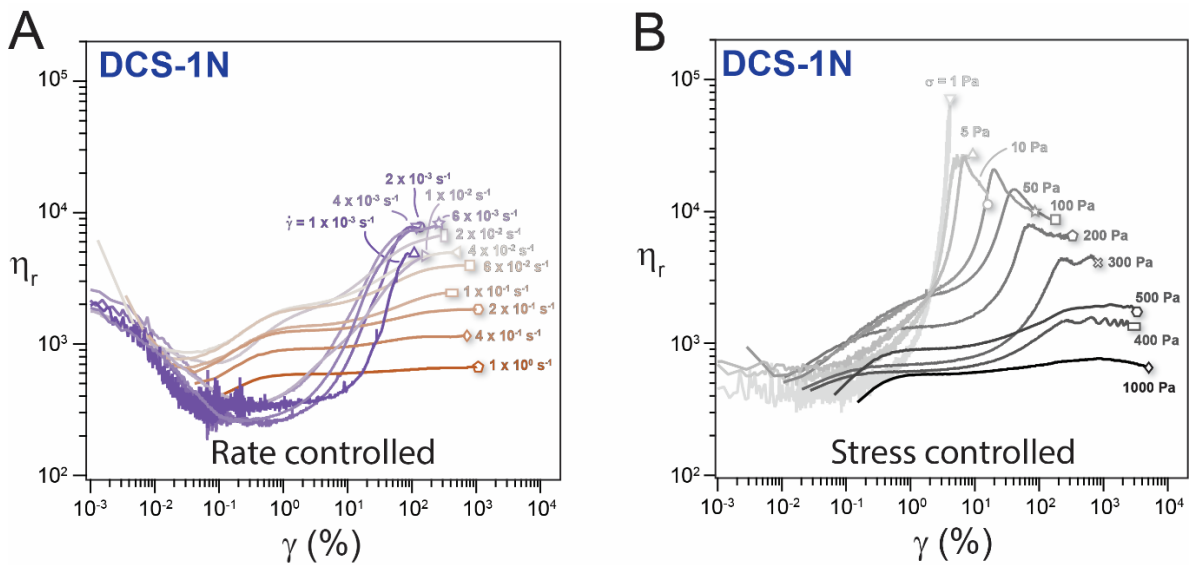

**Figure S11.** (*A*) Constant shear rate and measurements of **DCS-1N** at 25 °C. Reduced viscosity (η_r) increases with strain (γ) and plateaus to a high γ viscosity value. These high γ values are plotted in Figure 3A as the 25 °C steady state for **DCS-1N**. (*B*) Constant shear stress (*creep*) measurements also showing a viscosity bifurcation for σ ≤ 1 Pa. It is currently unknown why this system exhibits stress overshoots for σ ≥ 10 Pa.

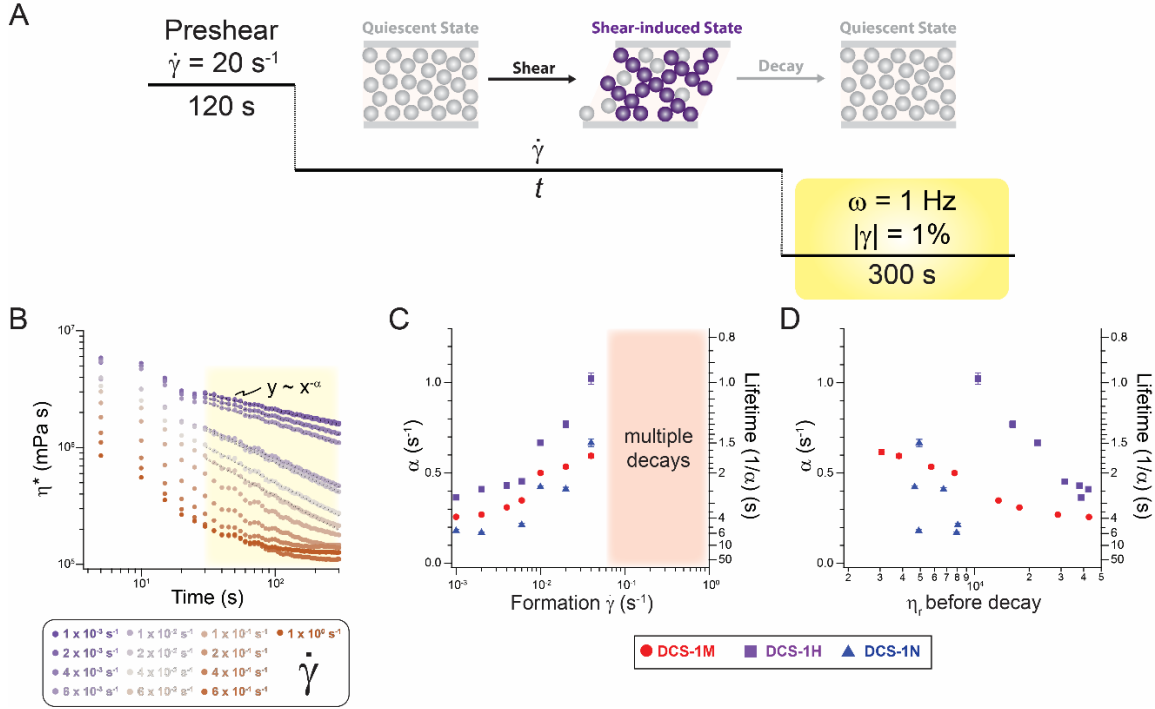

**Figure S12.** Oscillatory decay of the high viscosity antithixotropic state. (*A*) Experimental protocol: after an initial preshear, a constant shear rate was applied as shown in Figure 2 for **DCS-1M** and Figures S10-11 for **DCS-1H** and **DCS-1N**, followed by 300 s of oscillation at ω = 1 Hz and |γ| = 1%. (*B*) Representative decay of η* for **DCS-1M** networks constructed at different shear rates. (*C*) Power law exponent extracted from $t = 30 - 300$ s showing that structures formed at lower shear rates decay more slowly. (*D*) Power law exponent plotted against the particle network viscosity prior to shear cessation ($\eta_r$). Error bars are the standard error associated with the power law fitting.

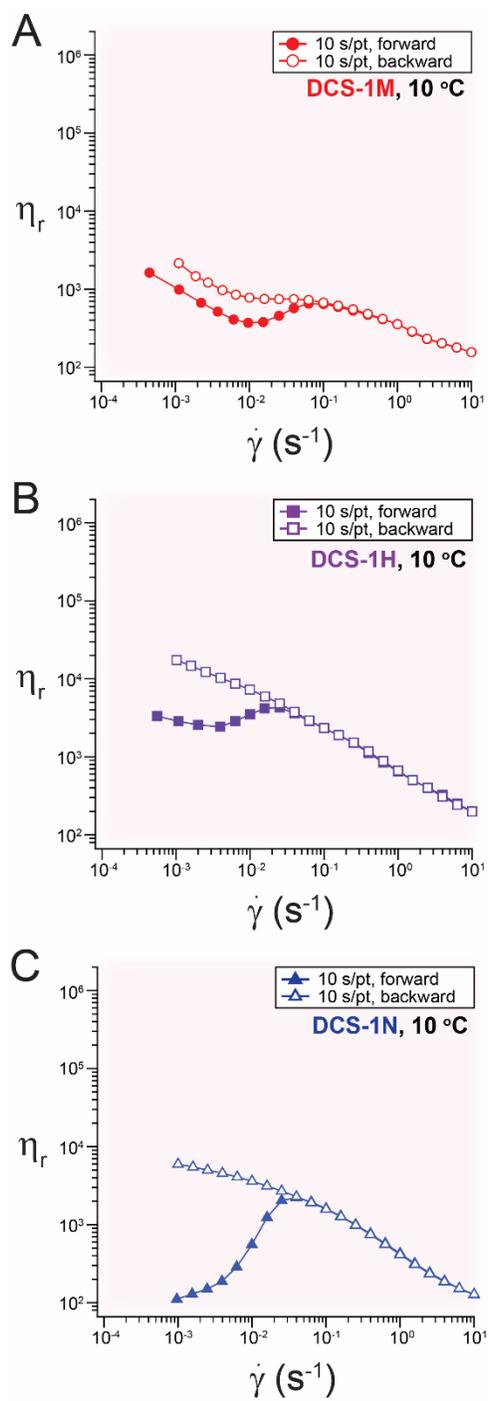

**Figure S13.** Hysteresis loops shows antithixotropy for all **DCSs** at 10 °C.

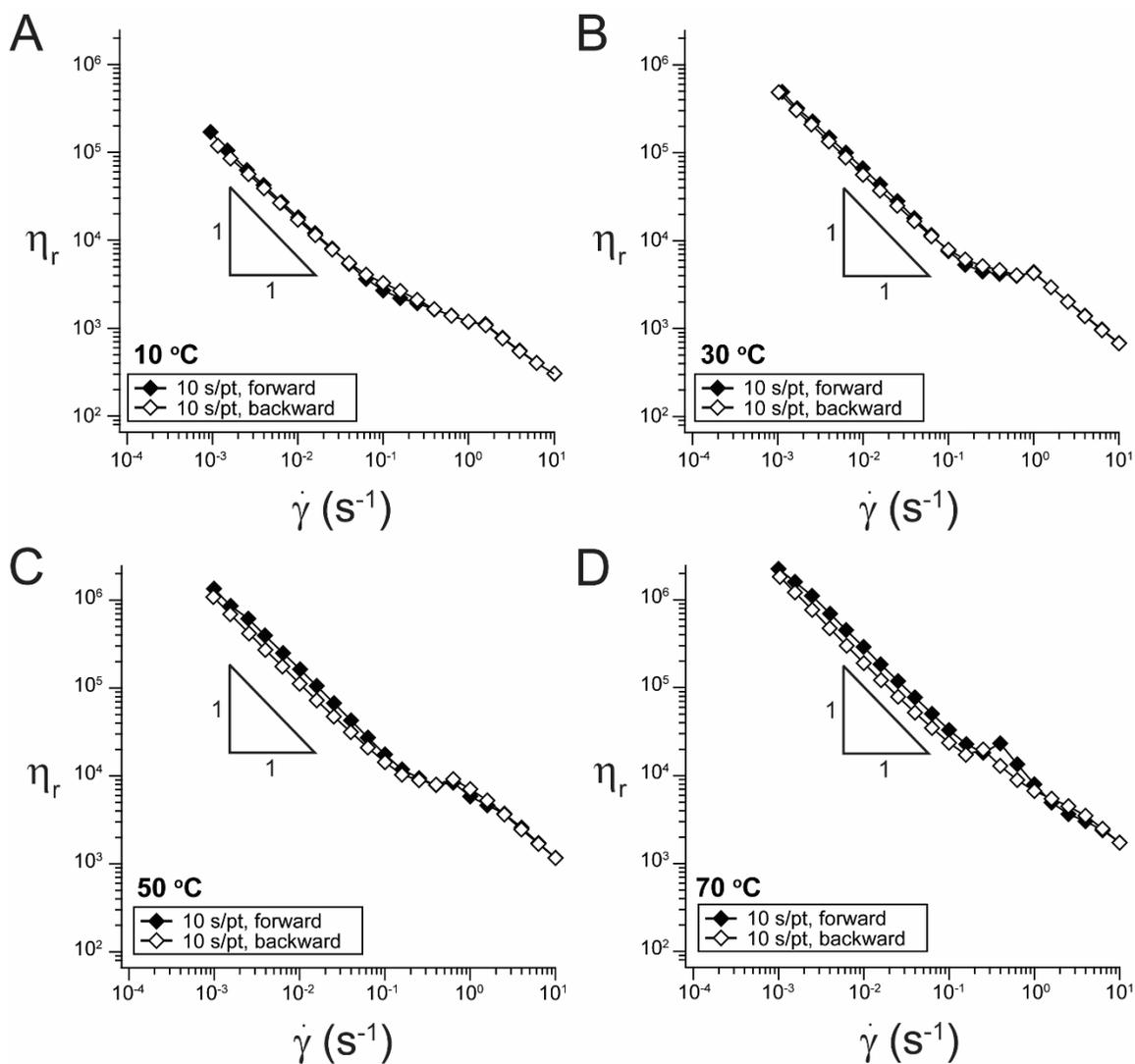

**Figure S14.** Temperature dependent shear rate ramps for **NCS-OH** showing shear thinning that is either reversible or with a small amount of thixotropy over the entire temperature range.

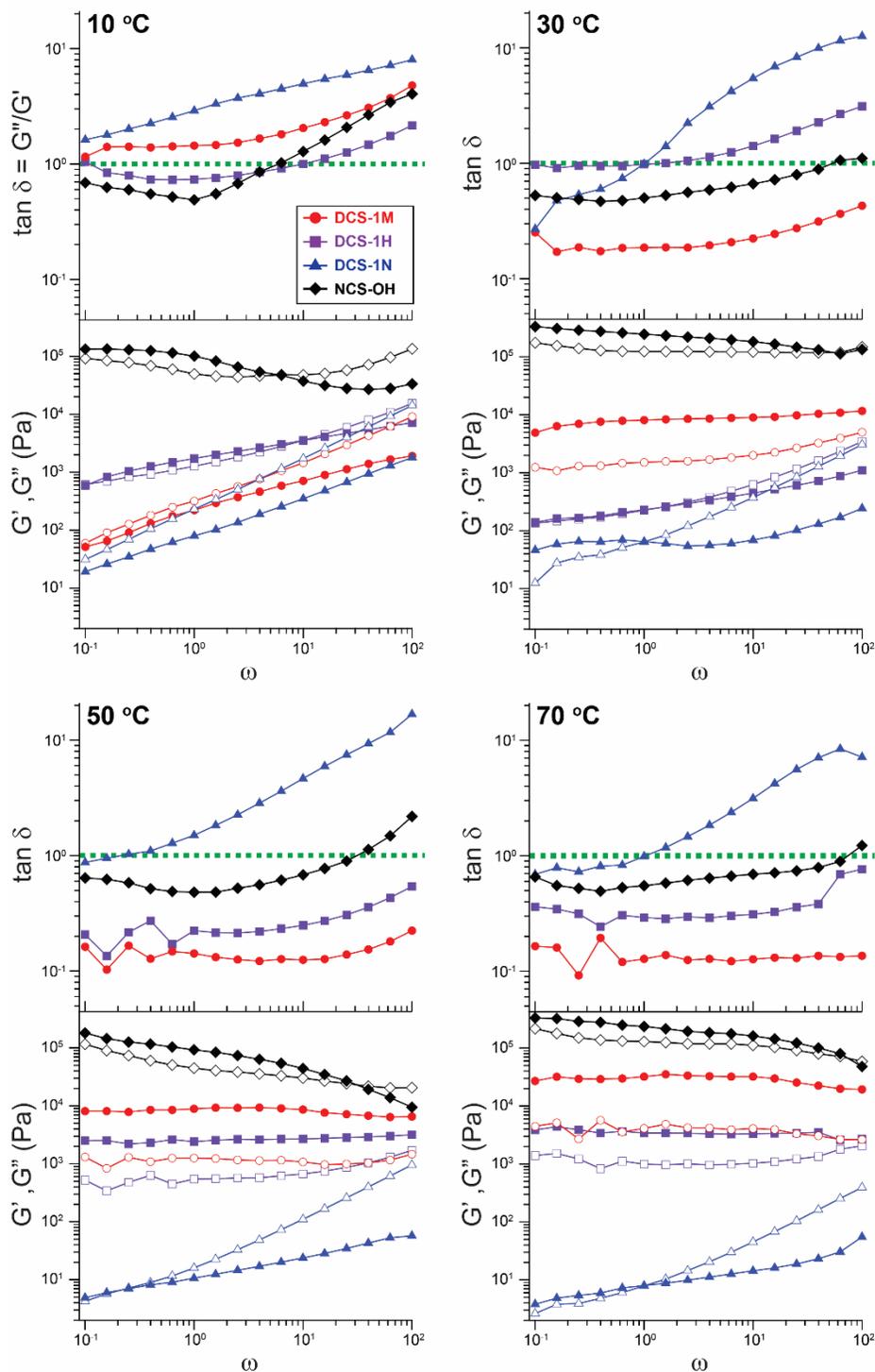

**Figure S15.** Temperature dependent small amplitude oscillatory shear (SAOS) measurements of **DCSs** and **NCS-OH**. Measurements conducted within the linear viscoelastic regime with |γ| = 0.5%. Heating **DCS-1M** shows a transition from liquid-like (tan δ > 1) at 10 °C to solid-like (tan δ < 1) ≥ 30 °C. This transition occurs between 40 °C and 50 °C for **DCS-1H**, whereas **DCS-1N** is liquid-like or viscoelastic (tan δ crosses 1 within the frequency range) at all temperatures.

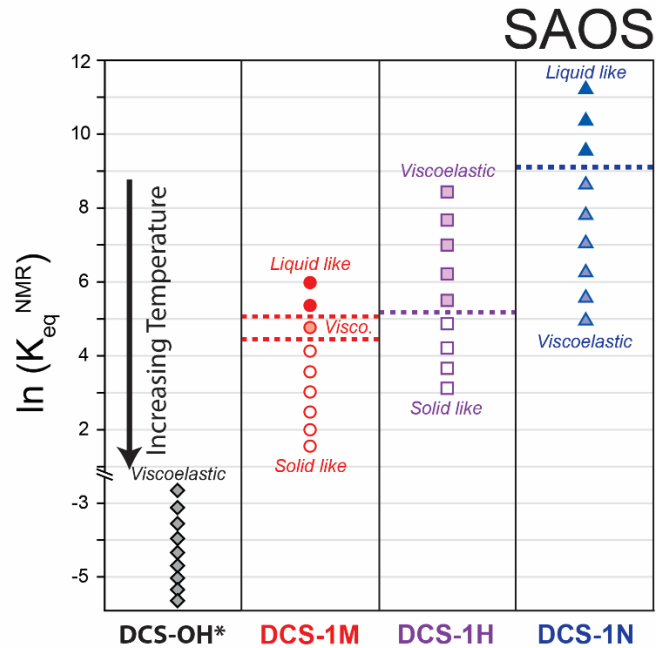

**Figure S16.** SAOS rheological state diagram as a function of $K_{eq}^{NMR}$ for 0 – 80 °C for each system from the data shown in Figure S15. This indicates the quiescent state of the suspensions, whereas the rheological state diagram shown in Figure 3B of the main text is for non-linear steady shear rheology.

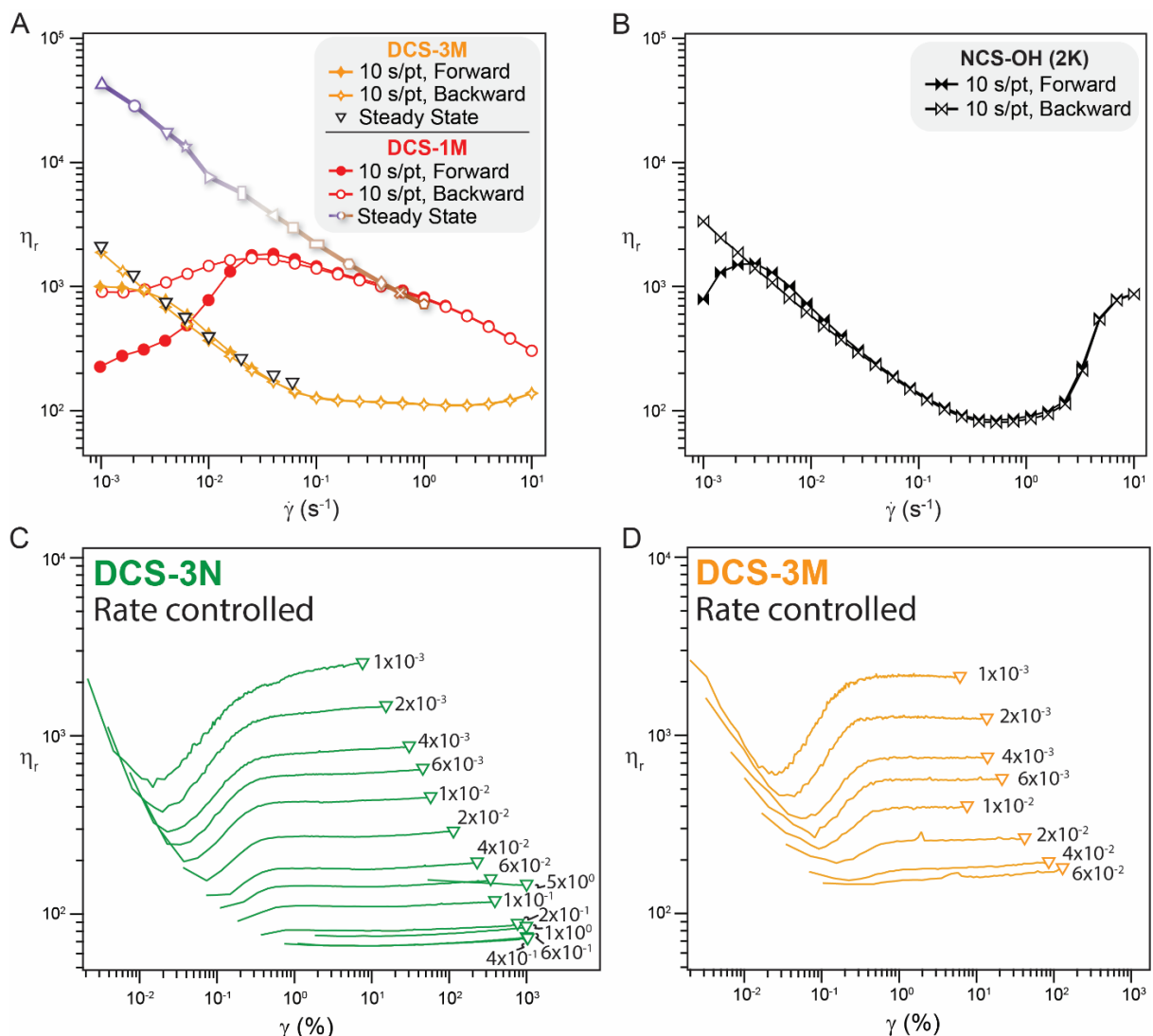

**Figure S17**. (*A*) Comparing **DCSs** with monotopic tMA **3M** and ditopic **1M**. Reduced viscosity ($\eta_r$) for a forward-backward shear rate ($\dot{\gamma}$) ramp reveals mostly reversible behavior for **DCS-3M**, with a backward shear rate ramp which matches the steady state viscosity. This behavior contrasts greatly with that of the **DCS-1M**, which shows pronounced hysteresis. The steady state reduced viscosity for **DCS-3M** is also much lower than that of **DCS-1M** at a given shear rate. (*B*) Demonstrating that the slight hysteresis at low shear rates ($\sim 10^{-3}$ s$^{-1}$) seen for **DCS-3M** and **DCS-3N** is also observed for a non-covalent suspension prepared with dihydroxy poly(propylene glycol) with $M_n$ $\sim 2,000$ g/mol **NCS-OH (2K)**. This feature is likely related to an inhomogeneous flow profile at short times.[10] (*C*) Constant shear rate measurements for **DCS-3N** showing that $\eta_r$ increases with strain ($\gamma$) and plateaus to a high $\gamma$ viscosity value. **DCS-3N** requires far less strain to equilibrate compared to **DCS-1N** (Figure S11). The high $\gamma$ values for **DCS-3N** are plotted in Figure 4B as the steady state values. (*D*) Constant shear rate measurements for **DCS-3M**, which also equilibrate at much lower strain values than **DCS-1M**. All data shown are for 25 °C.